\documentclass[11pt]{article} 
\usepackage{a4,multicol,graphicx,url,afterpage,float,rotating}
\usepackage[varg]{txfonts}
\usepackage{natbib}
\usepackage[utf8]{inputenc}
\usepackage{times}
\usepackage{microtype}
\usepackage[affil-it]{authblk}
\usepackage[margin={3cm,3cm}]{geometry}
\bibpunct[, ]{(}{)}{;}{a}{}{,}
  \usepackage{xcolor}
  \definecolor{teal}{RGB}{0,128,128}
  \usepackage[pdftitle={GMRT Technical Memo: Constraining Direction-Dependent Instrumental Polarisation},
              pdfauthor={Jamie Farnes, MRAO, Cambridge, UK},
              bookmarks=false,
              colorlinks=false,
              linkcolor=violet,
              citecolor=teal,
              urlcolor=magenta]{hyperref}

\restylefloat{figure}

\newcommand{\captionfonts}{\small}
\makeatletter  
\long\def\@makecaption#1#2{%
  \vskip\abovecaptionskip
  \sbox\@tempboxa{{\captionfonts #1: #2}}%
  \ifdim \wd\@tempboxa >\hsize
    {\captionfonts #1: #2\par}
  \else
    \hbox to\hsize{\hfil\box\@tempboxa\hfil}%
  \fi
  \vskip\belowcaptionskip}
\makeatother   



\title{Constraining Direction-Dependent Instrumental Polarisation: A New Technique for Polarisation Angle Calibration\\\vspace{0.3cm}\Large{\emph{National Centre for Radio Astrophysics (NCRA) Technical Reports}}\\\vspace{0.2cm}\large{v1.0: May, 2012; v2.0: July, 2014}}
\author{J. S. Farnes\thanks{jamie.farnes@sydney.edu.au}}
\affil{\large{Cavendish Laboratory, University of Cambridge, 19 J. J. Thomson Avenue, CB3 0HE, UK.
\\Sydney Institute for Astronomy, School of Physics, University of Sydney, NSW 2006, Australia.}}
\date{}

\begin{document}
\newpage
\pagenumbering{roman}
\maketitle
\begin{abstract}
Direction-dependent instrumental polarisation introduces wide-field polarimetric aberrations and limits the dynamic range of low-frequency interferometric images. We therefore provide a detailed two-dimensional analysis of the Giant Metrewave Radio Telescope (GMRT) primary beam in full-Stokes at 325~MHz and 610~MHz. We find that the directional dependence is essentially independent of the feed and is dominated by the curvature of the dishes reflecting mesh. The developed beam models are used to reduce wide-field instrumental polarisation in 610~MHz observations by subtracting the expected response from the $uv$-data itself. Furthermore, a new technique for polarisation angle calibration is presented that allows for calibration using an \emph{unpolarised} source and therefore can be implemented at arbitrarily low observational frequencies. This technique has the advantage that it calibrates the polarisation angle independently of ionospheric Faraday rotation and source variability. It also removes the need for known polarised sources on the sky -- which are scarce at low frequencies. We use the technique to retrieve the Rotation Measure of pulsar B1937+21 at 325~MHz, finding it to be consistent with previous independent measurements. An extended version of this method may be useful for verifying the calibration of other interferometers intended for polarimetric surveys.
\end{abstract}

\pagenumbering{arabic}
\section{Introduction}
\label{intro}
Polarised radio emission is fundamentally related to the presence of ordered magnetic fields, and observations of the polarisation properties of non-thermal radio emission are arguably the best way of studying quasi-regular fields and their structure. Polarimetric observations have recently become available at the Giant Metrewave Radio Telescope (GMRT). It is important to be able to complete a polarisation calibration that reliably removes instrumental effects, so that the polarisation mode may be used for scientific purposes. There are two essential steps in the initial polarisation calibration for an alt-az mounted interferometer with circular feeds, such as the GMRT:
\begin{enumerate}
\item Firstly, the orthogonal polarisations of antenna feeds are never perfectly isolated, and corrections must be made for this polarisation `leakage' between \(R\) and \(L\). This is typically carried out by observing an unresolved point source at the phase-centre over a range of parallactic angles, allowing for the separation of instrumental and source polarisation.
\item When solving for the instrumental leakage, the absolute value of the phase offset between \(R\) and \(L\) is left as an unconstrained degree of freedom. This typically needs to be corrected by observing a source with known electric vector polarisation angle (EVPA). The calibrators 3C138 and 3C286 are generally used for this purpose at higher frequencies.
\end{enumerate}

These two steps provide an `on-axis' polarisation calibration, and reduce residual instrumental polarisation to a minimum at the phase-centre. This calibration can take significant computing time at the GMRT as the leakages vary rapidly across the band, so the calibration must be carried out on each spectral channel individually. 

Any spurious instrumental polarisation results in leakage from Stokes \(I\) into the \(Q\) and \(U\) images. Following on-axis calibration at the GMRT, a residual instrumental polarisation of \(\le0.25\)\% can typically be obtained at the phase-centre \citep{Farnesthesis}. However, the response of an interferometer varies across the primary beam, and wide-field polarimetry requires calibration of these `direction-dependent' or `off-axis' instrumental effects. Similar to the case of on-axis calibration, the `polarisation beam' manifests itself with flux in total intensity leaking into the polarisation images. These polarimetric aberrations generally result in an increase in the observed fractional polarisation and also alter the absolute EVPA of sources -- with the effect typically becoming more pronounced with increasing distance from the phase-centre within the main lobe of the primary beam. Direction-dependent effects therefore limit the dynamic range of low-frequency interferometric images and restrict the region of the primary beam that is useful for scientific measurement.

The fundamental cause of direction-dependent instrumental linear polarisation is typically the curved reflecting surface, which slightly changes the direction of an incident electric vector upon reflection. For on-axis sources, these polarimetric aberrations cancel out. For off-axis sources, the path length to the source from different parts of the reflecting surface are not all equal. These distortions increase with curvature \citep{2002ASPC.proc..278S,1996aspo.book.....T}. The general equations to compute the off-axis instrumental polarisation beam are given by \citet{caretti2004}. From a more observational perspective, and following \citet{HeilesAreciboMemo}, there are two kinds of beam polarisation that are theoretically expected:
\begin{enumerate}
\item Beam squint: this occurs when the two circular polarisations point in different directions by a certain angle. Beam squint tends to produce a `two-lobed' pattern, one positive and one negative.
\item Beam squash: this occurs when the two linear polarisations have different beamwidths by a certain amount. Beam squash tends to produce a `four-lobed' pattern, in which two lobes on opposite sides of the beam centre have the same sign and two lobes rotated by \(90^{\circ}\) have the opposite sign. This quadrupolar pattern tends to give rise to instrumental linear polarisation that is oriented radially with respect to the phase-centre.
\end{enumerate}

A classical model of beam squint/squash in prime-focus feeds suggests that beam squash can arise in \(Q\)/\(U\) due to two effects: the interaction of the linearly polarised vectors with the curvature of the reflecting surface, and the difference between the two polarisations in illumination of the primary surface (which occurs from pseudo-waveguide modes in a feed). Similarly, beam squint should arise in \(V\) due to the feed not pointing directly at the vertex of the paraboloid. There is no anticipation of beam squint in \(Q\)/\(U\) or beam squash in \(V\) \citep{HeilesGBTMemo}. Knowledge of the pattern of beam squint and beam squash across the FOV, and the ability to correct for the resulting direction-dependent effects, is vital for mosaiced surveys and other observations where the science relies on wide-field polarimetric capabilities. 

In this report, an observational analysis of the GMRT beam is provided in full-polarisation at both 325~MHz and 610~MHz. The obtained polarisation beam model is then used to correct for the effects of direction-dependent instrumental polarisation at 610~MHz. Details of the holography observations used to obtain the beam constraints are described in Section \ref{chap6:observations}. The process used to retrieve the beam response from the holography data is detailed in Section \ref{chap6:rasterprocessing}, while two-dimensional maps of the full-Stokes beam at 610~MHz are presented in Section \ref{chap6:mappingresponse} along with an analysis of the 325~MHz beam. The obtained beam maps are then applied in order to correct the direction-dependent response. The theory behind these corrections and the impact on wide-field GMRT images is detailed in Section \ref{chap6:correctingresponse}. A new technique for polarisation angle calibration using an unpolarised calibrator is then detailed in Section \ref{chap6:anewtechniqueforpacalibration}. A discussion of the results is provided in Section \ref{chap6:discussion}.

\section{Observations}
\label{chap6:observations}
Observations were made with the GMRT, at both 325~MHz and 610~MHz using the software backend. All of the data used in this report are summarised in Table \ref{tab:observations}.
\begin{table}[ht]
\caption{Details of the observations.} 
\centering 
\begin{tabular}{c c c c c c c} 
\hline\hline 
Date & Obs. Code & Frequency /MHz & Bandwidth /MHz & Integration Time /s\\ [0.5ex] 
\hline 
2010 Jan 10 & 17\_052 & 610 & 16 & 16\\ 
2011 Apr 26 & TST0570 & 610 & 16 & 8\\ 
2011 Sep 09 & TST0620 & 325 & 32 & 16\\[1ex] 
\hline 
\end{tabular}
\label{tab:observations} 
\end{table}

Observation 17\_052 was a conventional full-track of the nearby galaxy M51 in full-polar mode, an analysis of which is fully presented in \citet{farnesMexico}. Standard calibration procedures for GMRT data were used for observation 17\_052, with the EVPA calibration applied using 3C286 which is assumed to have a rotation measure (RM) of \(-1.2\)~rad~m\(^{-2}\) and an intrinsic PA of \(33^{\circ}\) \citep{2009PhDT.........3G,Farnesthesis}. The observations TST0570 and TST0620 were also flagged and calibrated using similar techniques. 

The test observations TST0570 and TST0620 used for analysis of the GMRT full-Stokes beam were taken in modified holography mode. Two reference antennas remained fixed on an unpolarised calibrator while the remaining antennas were slewed in azimuth and/or elevation to a coordinate such that the calibrator source was observed offset from the phase-centre. In all cases, the two reference antennas were excluded from the data analysis so that only the offset antennas were used -- allowing the power pattern\footnote[1]{The power pattern is `superimposed' onto interferometric observations of the sky, and is therefore more fundamental for studying the effect that the beam has on measured source properties.} to be directly retrieved, rather than the voltage pattern.

\begin{figure}
\centering
    \includegraphics[height=93mm,clip=true,trim=0cm 0cm 0cm 0cm]{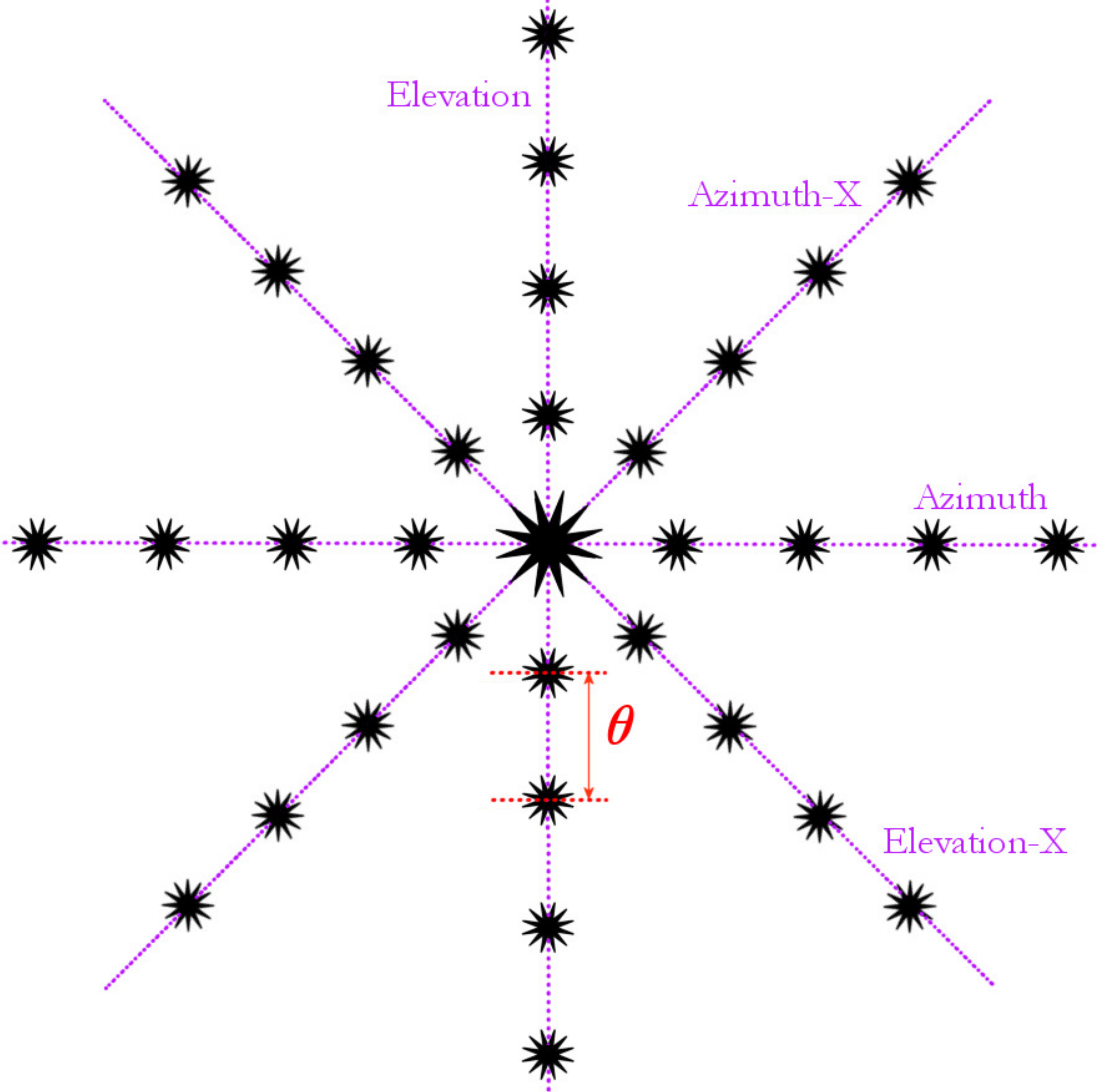}
  \caption{\small{The grid used to observe the off-axis raster scans. Each axis is labelled on the end that is defined as positive. Note that the Azimuth-X and Elevation-X axes were only observed at 610~MHz. For the 610~MHz observation, \(\theta = 10^{\prime}\), with 9 raster scans along each axis and a maximum offset of \(40^{\prime}\). The grid for the 325~MHz observation is similar, but uses \(\theta = 20^{\prime}\) to compensate for the larger beam, with a maximum offset of \(80^{\prime}\). The FWHM of the primary beam, as measured at the observatory, is \(85.2^{\prime}\) at \(325\)~MHz and \(44.4^{\prime}\) at \(610\)~MHz \citep{farnesMNRAS}.}}
  \label{offaxisgrid}
  \vspace{0pt}
\end{figure}

A number of offsets were used, such that a grid of `raster scans' were observed. The grid used for observing these off-axis raster scans is shown in Fig.~\ref{offaxisgrid}. Note that all four axes were only observed at 610~MHz, and the 325~MHz observations excluded the diagonal axes Azimuth-X and Elevation-X. Each axis of the observing grid included an on-axis scan of the holography source. These on-axis scans were used to solve for the complex gains of each antenna, with the solutions being interpolated across the other raster scans -- essentially using the on-axis scans for phase calibration. The effects of instrumental polarisation (and parallactic angle variation at the phase-centre) were removed from the data by solving for the leakages using all on-axis observations of the unpolarised calibrator. 

A polarisation angle calibration was applied to obtain the \(Q\) and \(U\) beam in the calibrated antenna frame. The correction to the holography data was applied so that the raster scans of the derotated azimuth axis displayed an EVPA that was oriented radially with respect to the phase-centre. Evidence to justify this assumption is provided in \citep{farnesMexico}.

For the sake of completeness, constraints on the Stokes \(V\) beam are obtained in Section \ref{chap6:rasterprocessing}, but no further attempt is made to correct for this response. Unless otherwise specified, the term `polarised' refers to linear polarisation throughout this report. The sources used for flux and polarisation leakage calibration, along with the unpolarised holography source observed for the off-axis raster scans are summarised in Table \ref{tab:sources}.
\begin{table*}[ht]
\caption{Details of the sources used for each observation.} 
\centering 
\begin{tabular}{c c c c c c} 
\hline\hline 
Obs. Code & Frequency /MHz & Flux \& Leakage cal. & Holography source\\ [0.5ex] 
\hline 
TST0570 & 610 & 3C147 & 3C147 \\ 
TST0620 & 325 & 3C48 \& 3C286 & 3C48\\[1ex] 
\hline 
\end{tabular}
\label{tab:sources} 
\end{table*} 

\section{Retrieving the full-Stokes beam from holography rasters}
\label{chap6:rasterprocessing}
The intention is to measure the direction-dependent instrumental polarisation response so that corrections can be enacted to the \(uv\)-data. Optimal corrections may ultimately require individual beam models for each antenna, however the computational expense would be prohibitive. As such, the effect of the polarisation beam is best represented here by the response of the \emph{average} antenna.

The \textsc{aips} task \textsc{listr} was used to output the average amplitude and phase of all visibilities, for each holography raster in all four cross-correlations. For an ideal, calibrated interferometer with circular feeds, the cross-correlations are complex quantities defined as,
\begin{equation}
RR = \mathcal{A}(RR)e^{i\psi_{RR}} = I+V \\,
\label{RRcomplex}
\end{equation}
\begin{equation}
LL = \mathcal{A}(LL)e^{i\psi_{LL}} = I-V \\,
\label{LLcomplex}
\end{equation}
\begin{equation}
RL = \mathcal{A}(RL)e^{i\psi_{RL}} = Q+iU \\,
\label{RLcomplex}
\end{equation}
\begin{equation}
LR = \mathcal{A}(LR)e^{i\psi_{LR}} = Q-iU \\,
\label{LRcomplex}
\end{equation}
where \(\mathcal{A}(jk)\) and \(\psi_{jk}\) are the amplitude and phase respectively. For a calibrated point-source, \(\psi_{RR}=\psi_{LL}=0\), so the phase terms of equations \ref{RRcomplex} and \ref{LLcomplex} can be neglected. However, the terms \(\psi_{RL}\) and \(\psi_{LR}\) cause mixing of \(Q\) and \(U\) and cannot be ignored.

By applying Euler's formula to equations \ref{RLcomplex} and \ref{LRcomplex}, and then substituting into \(RL=LR^{\ast}\) yields,
\begin{equation}
\left\{\mathcal{A}(RL)\cos\psi_{RL} - \mathcal{A}(LR)\cos\psi_{LR}\right\} + i\left\{\mathcal{A}(RL)\sin\psi_{RL} + \mathcal{A}(LR)\sin\psi_{LR}\right\} = 0 \\,
\label{aftereuler}
\end{equation}
which has the solution,
\begin{equation}
\mathcal{A}(RL)\cos\psi_{RL} - \mathcal{A}(LR)\cos\psi_{LR} = 0 \\,
\label{realequalszero}
\end{equation}
\begin{equation}
\mathcal{A}(RL)\sin\psi_{RL} + \mathcal{A}(LR)\sin\psi_{LR} = 0 \\.
\label{imagequalszero}
\end{equation}
These solutions can be used to show that for real data, normalised for the effects of the Stokes \(I\) primary beam, the fractional polarimetric beam response at a given raster is,
\begin{equation}
\frac{Q}{I} = \frac{\Re(RL+LR)}{RR+LL} \\,
\label{Qrealdata}
\end{equation}
\begin{equation}
\frac{U}{I} = \frac{\Im(RL-LR)}{RR+LL} \\,
\label{Urealdata}
\end{equation}
\begin{equation}
\frac{V}{I} = \frac{RR-LL}{RR+LL} \\,
\label{Vrealdata}
\end{equation}
while the Stokes \(I\) response itself can be obtained from,
\begin{equation}
I = \frac{RR+LL}{2} \\.
\label{Irealdata}
\end{equation}

A code was written in \textsc{python} and used to calculate the polarisation beam response from the amplitude and phase output by \textsc{aips}, in conjunction with equations \ref{Qrealdata} to \ref{Irealdata}. This was done for each holographic raster in each of the individual 220 channels across the band. All of the holography sources are assumed to be unpolarised at these frequencies, so no correction for source polarisation was required.

\begin{figure}[htp]
\centering
    \resizebox{57mm}{!}{\includegraphics[height=60mm,angle=-90,clip=true,trim=0cm 0cm 0cm 0cm]{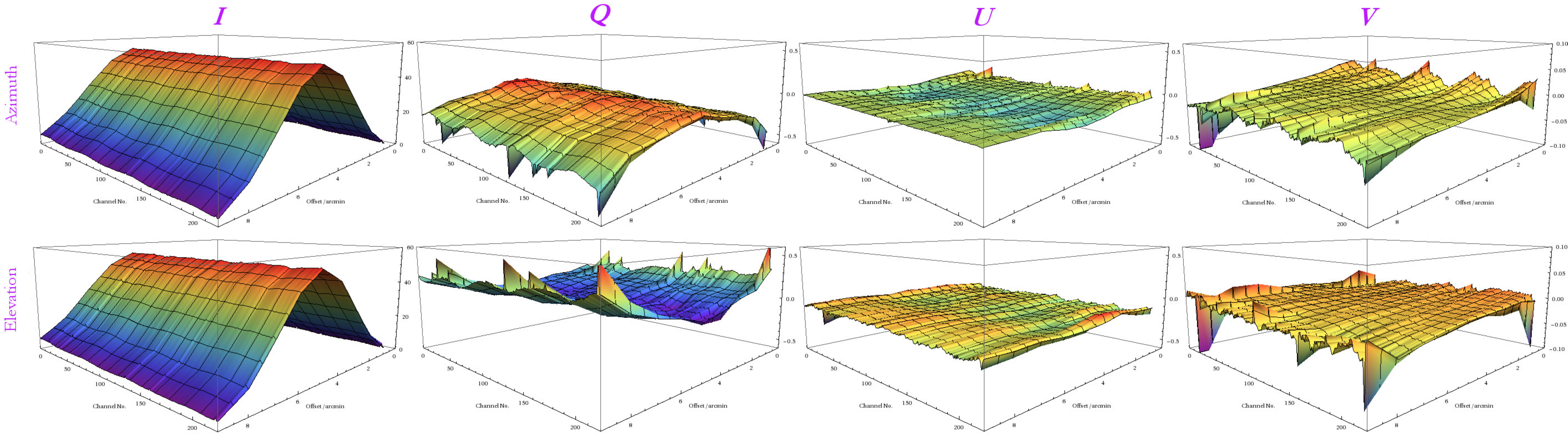}}
  \caption[Beam profiles at 325~MHz]{\small{The beam profiles across the band at 325~MHz. The \(x\)-axis displays the offset from the phase-centre, the \(y\)-axis displays the beam response and is in units of 0 to 60~Jy for the unnormalised Stokes \(I\) beam, and units of fractional polarisation for the other Stokes parameters -- the scales range from \(\pm60\)\% for \(Q\) and \(U\), and \(\pm10\)\% for \(V\). The \(z\)-axis displays the channel number across the band.}}
  \label{beamprofiles325}
  \vspace{0pt}
\end{figure}

\begin{figure}[htp]
\centering
    \resizebox{111mm}{!}{\includegraphics[height=115mm,angle=-90,clip=true,trim=0cm 0cm 0cm 0cm]{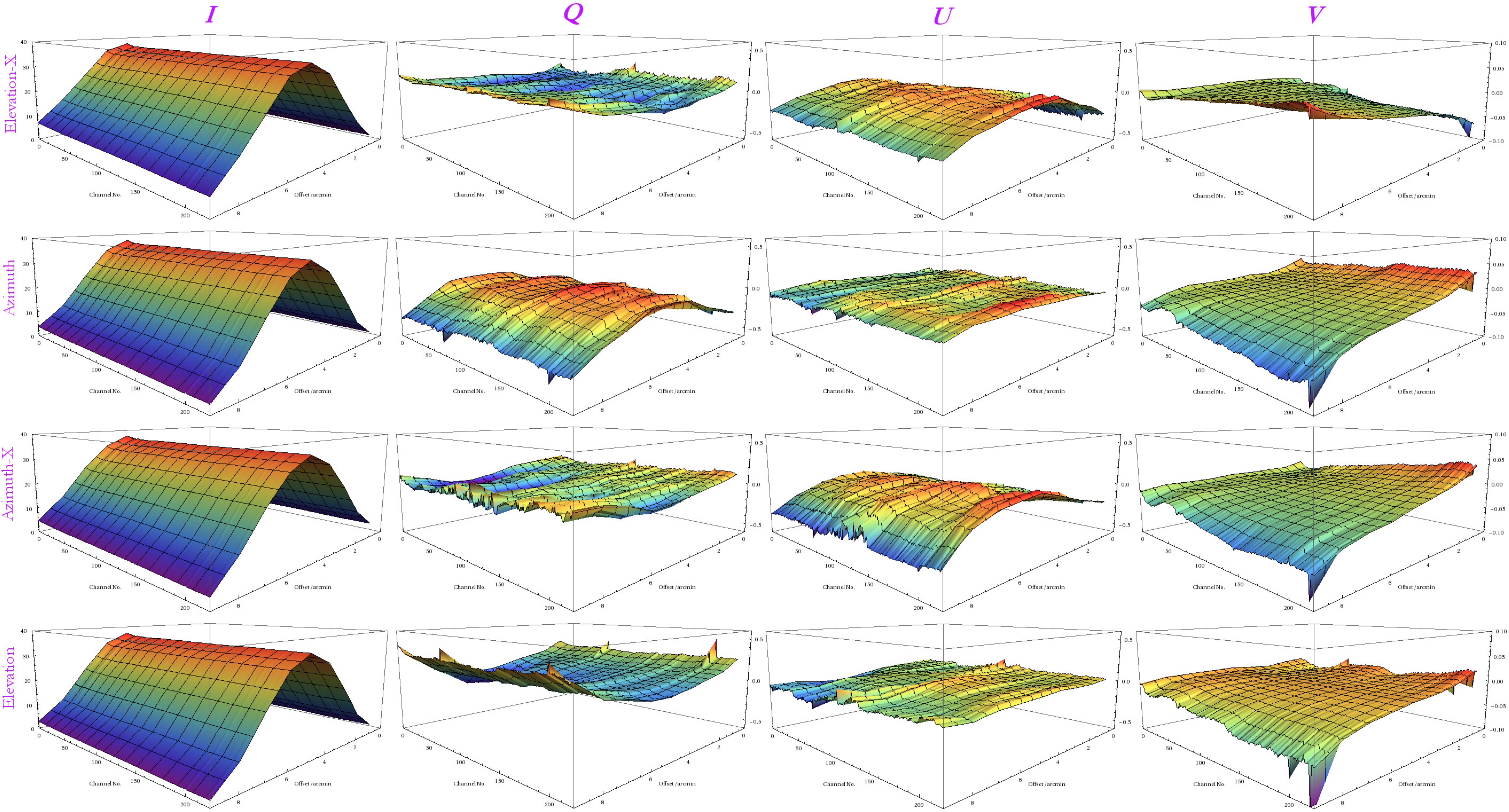}}
  \caption[Beam profiles at 610~MHz]{\small{The beam profiles across the band at 610~MHz. The \(x\)-axis displays the offset from the phase-centre, the \(y\)-axis displays the beam response and is in units of 0 to 40~Jy for the unnormalised Stokes \(I\) beam, and units of fractional polarisation for the other Stokes parameters -- the scales range from \(\pm60\)\% for \(Q\) and \(U\), and \(\pm10\)\% for \(V\). The \(z\)-axis displays the channel number across the band.}}
  \label{beamprofiles}
  \vspace{0pt}
\end{figure}

The instrumental polarisation and effects of rotation due to the parallactic angle, \(\chi\), have only been corrected at the phase-centre. These effects are actually direction-dependent and due to the GMRT's alt-az mount, any residual instrumental polarisation will display an EVPA that rotates with \(\chi\). All off-axis holographic raster scans had to be corrected for the effects of this \(\chi\)-dependent mixing of \(Q\) and \(U\). This erroneous complication of the earlier on-axis calibration has to be undone by derotating \(Q\) and \(U\), so that,
\begin{equation}
(Q^{\prime} + iU^{\prime}) = (Q + iU)e^{-2i\chi} \\.
\label{QUmixing}
\end{equation}
The derotated \(Q^{\prime}\) and \(U^{\prime}\) can then be used to form a final beam profile in full-Stokes. The obtained beam profiles in full-polarisation across the band are shown at 325~MHz in Fig.~\ref{beamprofiles325}, and at 610~MHz in Fig.~\ref{beamprofiles}.

\begin{figure}[p]
\centering
    \resizebox{140mm}{!}{\includegraphics[height=140mm,angle=0,clip=true,trim=0cm 0cm 0cm 0cm]{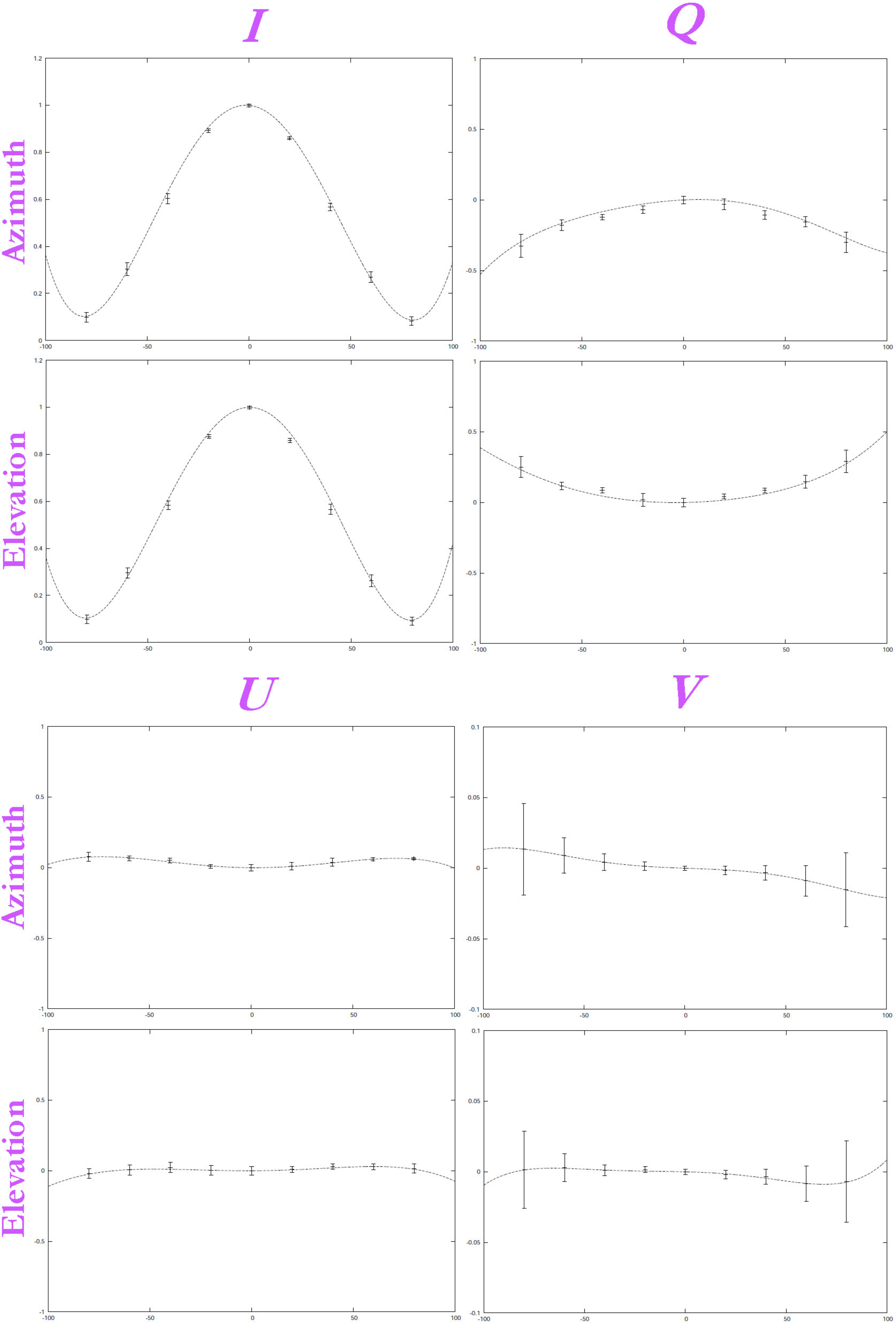}}
  \caption[Polynomial fits to the beam axes at 325~MHz]{\small{The 5th order polynomial fits to both beam axes for all Stokes parameters at 325~MHz, showing the normalised fractional response as a function of distance from the phase-centre. Stokes \(I\) has scales from 0 to 1.2, \(Q\) and \(U\) have scales from \(\pm1\), and \(V\) from \(\pm0.1\). Note that the data were sampled out to \(80^{\prime}\).}}
  \label{beamfits325}
  \vspace{0pt}
\end{figure}

\begin{figure}[p]
\centering
    \resizebox{140mm}{!}{\includegraphics[height=140mm,angle=0,clip=true,trim=0cm 0cm 0cm 0cm]{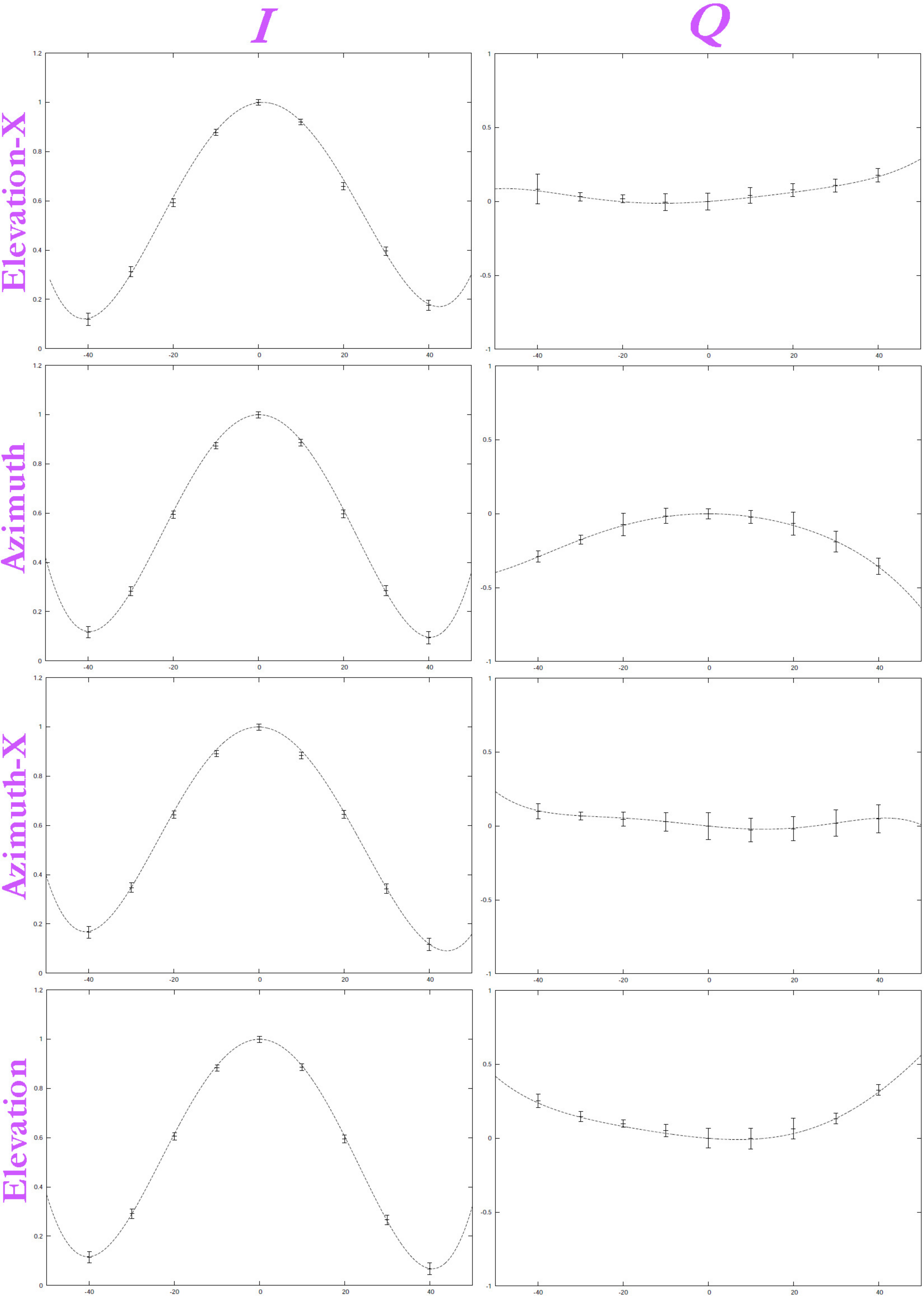}}
  \caption[Polynomial fits to I and Q of the beam axes at 610~MHz]{\small{The 5th order polynomial fits to each beam axis for Stokes parameters \(I\) and \(Q\) at 610~MHz, showing the normalised fractional response as a function of distance from the phase-centre. Stokes \(I\) has scales from 0 to 1.2, while \(Q\) has scales from \(\pm1\). The fits to \(U\) and \(V\) are shown in Fig.~\ref{beamfitsuv}. Note that the data were sampled out to \(40^{\prime}\).}}
  \label{beamfitsiq}
  \vspace{0pt}
\end{figure}

\begin{figure}[p]
\centering
    \resizebox{140mm}{!}{\includegraphics[height=140mm,angle=0,clip=true,trim=0cm 0cm 0cm 0cm]{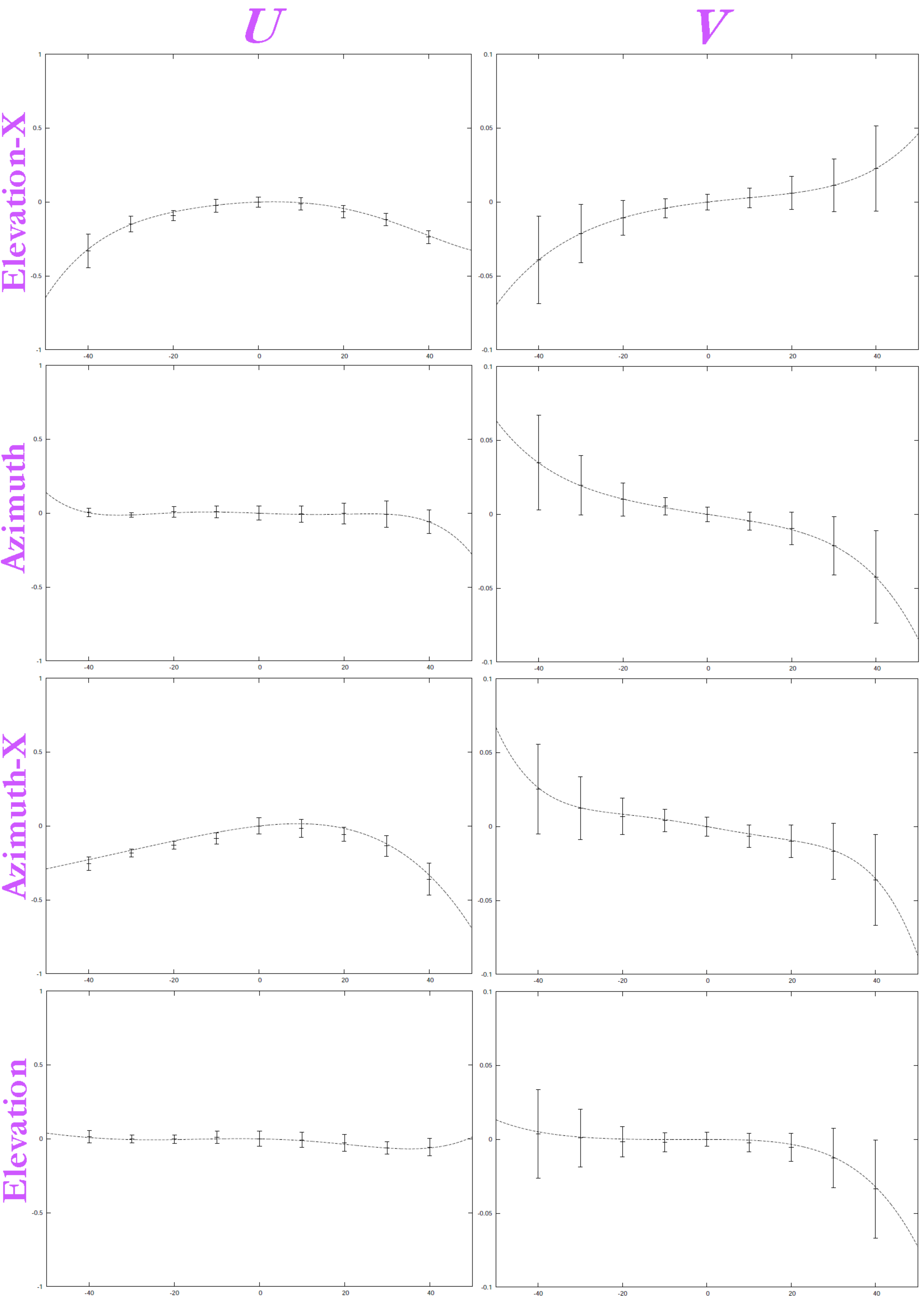}}
  \caption[Polynomial fits to U and V of the beam axes at 610~MHz]{\small{The 5th order polynomial fits to each beam axis for Stokes parameters \(U\) and \(V\) at 610~MHz, showing the normalised fractional response as a function of distance from the phase-centre. Stokes \(U\) has scales from \(\pm1\), and \(V\) from \(\pm0.1\). The fits to \(I\) and \(Q\) are shown in Fig.~\ref{beamfitsiq}. Note that the data were sampled out to \(40^{\prime}\).}}
  \label{beamfitsuv}
  \vspace{0pt}
\end{figure}

The mean response across the band was calculated for each holography raster, so that the beam was described by a single datapoint per raster -- creating a set of nine data per beam axis. Each axis was then fitted with a 5th order polynomial in Stokes \(I\), \(Q\), \(U\), and \(V\) using an ordinary least-squares fit. All of the one-dimensional slices through the beam and the corresponding fits are shown at 325~MHz in Fig.~\ref{beamfits325}, and at 610~MHz in Figs.~\ref{beamfitsiq} to \ref{beamfitsuv}. The general shape of each beam slice is well described by the fits.

Beyond the sampling limits of \(40^{\prime}\) or \(80^{\prime}\) at 610~MHz or 325~MHz respectively, the unconstrained polynomial fits deviate from the likely beam response. This has led to the introduction of artefacts in the beam maps at radii further away from the phase-centre than the sampling limit.

\subsection{FWHM of the Stokes I beam}
\label{chap6:beamFWHM}
Following polynomial fitting, the FWHM of the Stokes \(I\) beam axes at 610~MHz were found to be (\(46.5^{\prime}\pm0.8^{\prime}\)), (\(50.0^{\prime}\pm0.9^{\prime}\)), (\(46.5^{\prime}\pm0.8^{\prime}\)), and (\(49.8^{\prime}\pm0.8^{\prime}\)) in Azimuth, Azimuth-X, Elevation, and Elevation-X respectively. This suggests the main lobe of the Stokes \(I\) beam has a square--circle (or `squircle') shaped geometry at 610~MHz, with the FWHM of the beam increasing by \(\sim7.5\)\% along the diagonal axes.

The FWHM of the Stokes \(I\) beam at 325~MHz was found to be (\(92.3^{\prime}\pm1.2^{\prime}\)) and (\(93.1^{\prime}\pm1.2^{\prime}\)) in Azimuth and Elevation respectively.

The measured FWHM of the primary beam differs from the values determined at the observatory of \(85.2^{\prime}\) at \(325\)~MHz and \(44.4^{\prime}\) at \(610\)~MHz. This is likely a consequence of both the fitting procedure and expected variations in the Stokes \(I\) beam -- see Section \ref{chap6:discussion} for additional detail.

\section{Two-dimensional mapping of the full-Stokes beam}
\label{chap6:mappingresponse}
The polynomial fits from Section \ref{chap6:rasterprocessing} were used in order to create two-dimensional maps of the GMRT beam at 610~MHz. As the 325~MHz beam was not sampled along the diagonal axes, it was not possible to create a two-dimensional map at this frequency.

The polynomials describe the off-axis response for each Stokes parameter as a function of radius from the phase-centre, and so were used to tangentially interpolate and grid the beam response around the phase-centre at a given radius. This was done using cubic spline interpolation in order to smoothly interpolate the sparsely sampled data over the large tangential distances. Polynomial fitting may have given rise to significant Runge's phenomena.

Further consideration of the 610~MHz polarisation beam response along the diagonal axes suggests a significant problem with the Elevation-X axis. The direction-dependent response along this axis, as displayed in Figs.~\ref{beamfitsiq} to \ref{beamfitsuv}, appears similar to the Azimuth-X axis. Consequently, polarisation vectors along the Elevation-X axis were oriented tangentially about the phase-centre in the initial \(Q\) and \(U\) maps -- in disagreement with the results of RM Synthesis and the expected quadrupolar response of the GMRT polarisation beam, as discussed in \citep{farnesMexico}. 

\begin{figure}[tb]
\centering
    \resizebox{135mm}{!}{\includegraphics[height=145mm,angle=0,clip=true,trim=0cm 0cm 0cm 0cm]{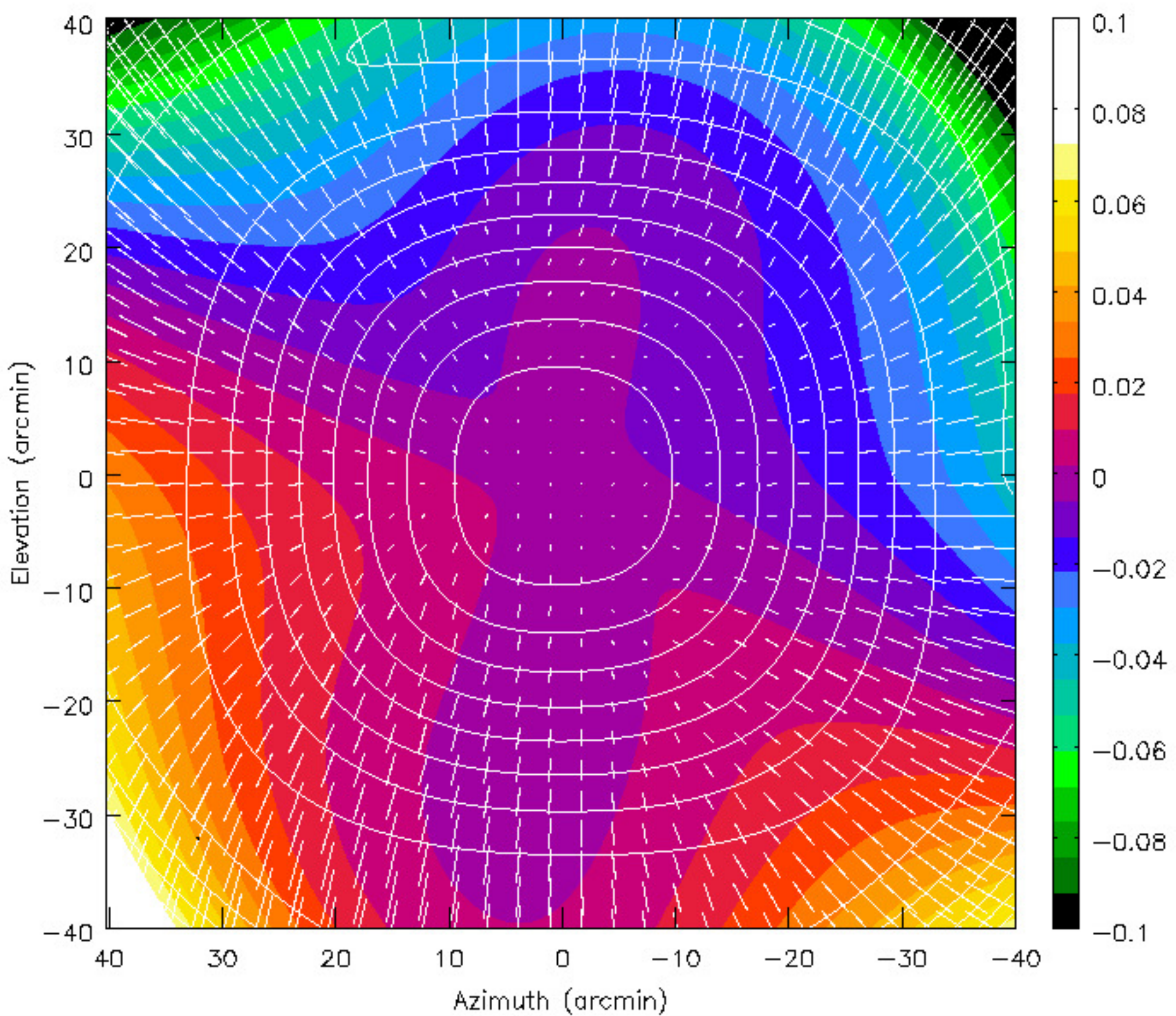}}
  \caption[The GMRT 610~MHz beam in full-polarisation]{\small{The main lobe of the GMRT primary beam in full-Stokes at 610~MHz. White contours show the Stokes \(I\) response with contour levels at 10, 20, 30, ..., 90\%. White polarisation vectors are proportional to the linearly polarised intensity and indicate the orientation of the linearly polarised response. The Stokes \(V\) response is shown by the pseudo-colour scale, which ranges from \(\pm10\)\% \citep{farnesMNRAS}.}}
  \label{allbeam}
  \vspace{-10pt}
\end{figure}

Attempts to correct for GMRT direction-dependent instrumental polarisation (see Section \ref{chap6:correctingresponse}) failed using this initial beam model, with the polarisation of all sources tending to increase. Images made in \textsc{aips} of the off-axis holography sources confirmed that the unusual beam response was a result of the observation itself. As a test, the orientation of polarisation vectors along the Elevation-X axis were rotated so that they were forced to be oriented radially -- this rotation was equivalent to a change of sign for both \(Q\) and \(U\). The resulting maps are consistent with the response of the GMRT inferred using RM Synthesis, and appear to adequately correct direction-dependent effects. No explanation for this anomaly has been found, and it is crucial that this be checked by further observation.\footnote[2]{Update (2014): Follow up observations have confirmed that there was a bug in the software used to calculate the coordinates of the Elevation-X axis, resulting in the Azimuth-X axis being observed a second time.}

\begin{figure}[htp]
\centering
    \resizebox{110mm}{!}{\includegraphics[height=77mm,angle=0,clip=true,trim=0cm 0cm 0cm 0cm]{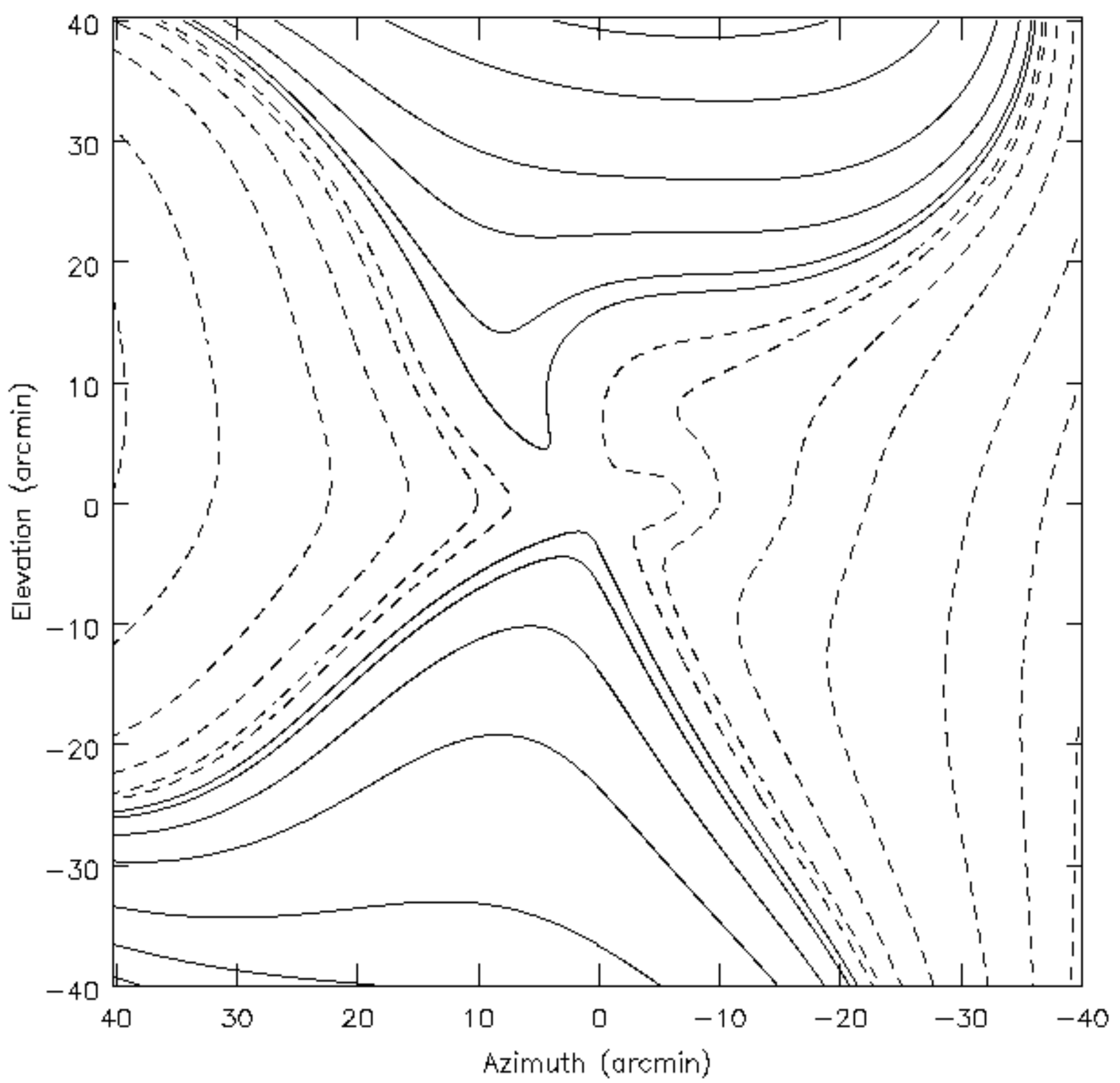}}  \\
    \resizebox{110mm}{!}{\includegraphics[height=77mm,angle=0,clip=true,trim=0cm 0cm 0cm 0cm]{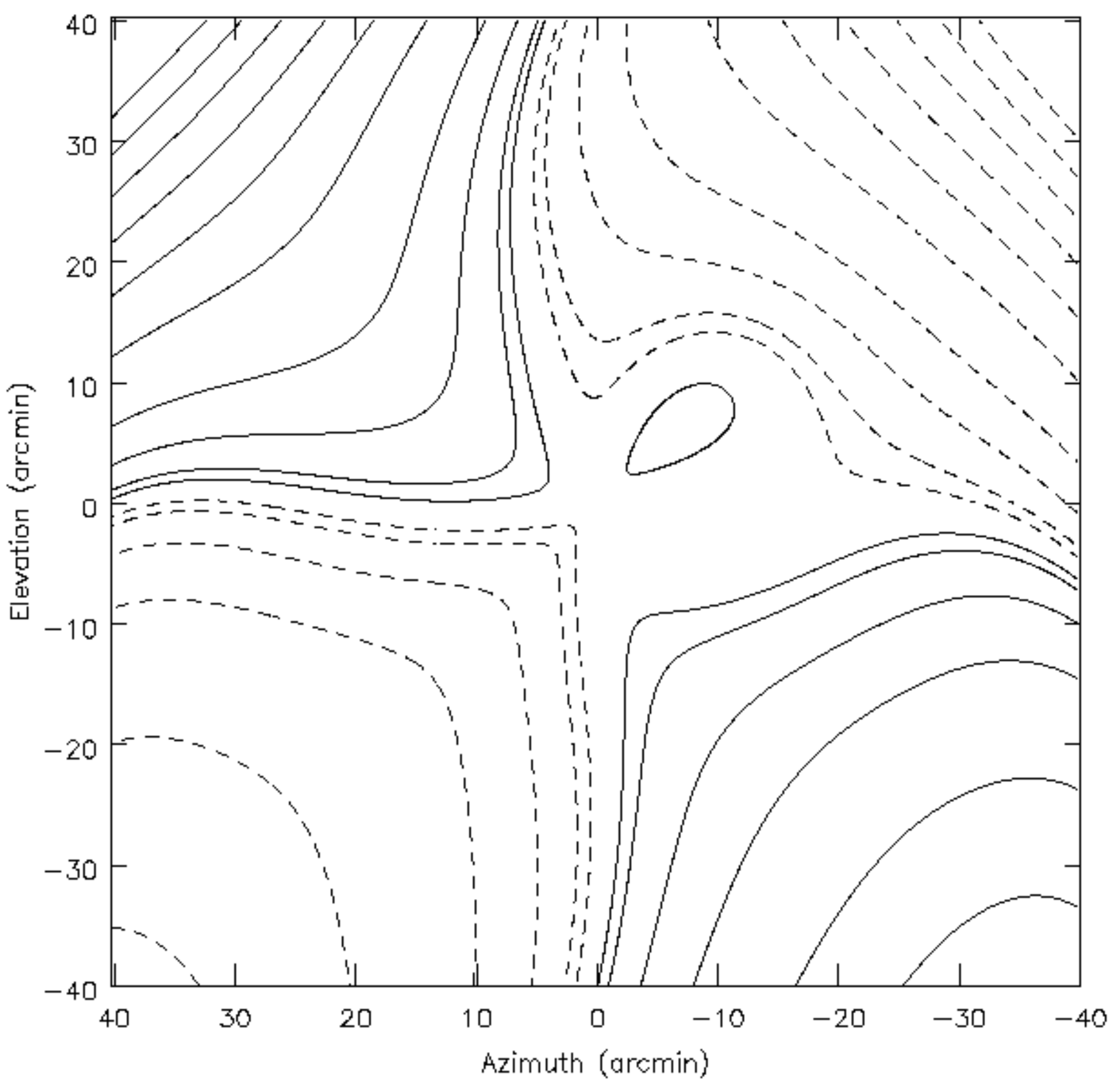}}
  \vspace{0pt} \\
  \caption[The GMRT Stokes Q and U 610~MHz beams]{\small{The Stokes \(Q\) (top) and \(U\) (bottom) fractional instrumental polarisation beam of the main lobe of the GMRT at 610~MHz. Contour levels are at \(\pm1\), 2, 5, 10, 20, 30, ..., 70\% \citep{farnesMNRAS}.}}
  \label{QUbeam}
  \vspace{0pt}
\end{figure}

The final two-dimensional beam at 610~MHz, interpolated under the assumption that the polarisation vectors are oriented radially along all axes, is shown in Figs.~ \ref{allbeam} to \ref{QUbeam}.

\section{Correcting the Wide-field Response}
\label{chap6:correctingresponse}
The wide-field response of a radio interferometer is modulated by the full polarisation beam. In order to perform accurate wide-field polarimetry, it is essential to be able to correct for these direction-dependent beam effects. No attempt was made to correct for the effects of spurious circular polarisation. 

For linear polarisation observations with alt-az mounted circular feeds, the introduced aberrations are a consequence of three effects. For simplicity, I first consider a `snapshot' observation, such that the data can be considered as of constant parallactic angle. For this case, polarimetric aberrations are introduced by:
\begin{enumerate}
\item The \(Q\) and \(U\) beam being rotated in the sky-plane by an amount equal to the parallactic angle.
\item Mixing of the \(Q\)/\(U\) beam response as a consequence of the parallactic angle.
\item The resulting rotated and mixed \(Q\) and \(U\) fractional polarisation beams allow a fraction of the total intensity to leak into the \(Q\) and \(U\) images.
\end{enumerate}
For a single snapshot, it is possible to correct for these effects by first rotating the derived \(Q\) and \(U\) fractional polarisation beams, shown in Fig.~\ref{QUbeam}, by an angle equal to \(\chi\) about the phase-centre. The mixing of \(Q\) and \(U\) is introduced by the on-axis polarisation calibration and the associated correction for parallactic angle variation at the phase-centre -- causing any spurious polarisation to rotate in the complex-plane as a function of \(\chi\). Similar corrections have previously been applied in the image-plane to VLA data for snapshot observations \citep{CottonAIPSMemo}.

As most GMRT polarimetric observations are full-track observations, an image-plane based procedure is clearly not ideal. As discussed in \citet{CottonAIPSMemo}, it is possible to remove direction-dependent effects from full-track observations by independently correcting and imaging each snapshot of constant parallactic angle and averaging the resulting images. However, as \textsc{clean} is a non-linear algorithm, the average of separately imaged `chunks' of \(uv\)-data is not identical to a single image of averaged \(uv\)-data. This limitation has been overcome using a novel technique whereby the clean components from a Stokes \(I\) image are scaled by the previously mentioned \(\chi\)-dependent effects, and the response removed in the \(uv\)-plane.

Having interpolated the beam model derived in Section \ref{chap6:mappingresponse} to the same geometry as the Stokes \(I\) image, the direction-dependent response, \(R(x,y,\chi)\), for \(k\) clean components is described by,
\begin{equation}
R = \sum_{i=1}^{k} I_{i}(x,y) \left\{ Q(x^{\prime},y^{\prime}) + iU(x^{\prime},y^{\prime}) \right\} e^{2i\chi} \owns \left[
\begin{array}{c}
x^{\prime} \\
y^{\prime}
\end{array} \right]
=
\left[
\begin{array}{cc}
\cos\chi & \sin\chi \\
-\sin\chi & \cos\chi
\end{array} \right]
\left[
\begin{array}{c}
x \\
y
\end{array} \right]  \\,
\label{fullresponse}
\end{equation}
where \(I_{i}(x,y)\) is the Stokes \(I\) flux of the \(i\)-th clean component and \(Q(x^{\prime},y^{\prime})\) \& \(U(x^{\prime},y^{\prime})\) are the fractional polarisation response obtained from the beam model. The pixel coordinates \((x,y)\) and \((x^{\prime},y^{\prime})\) are integers as a consequence of pixel quantisation in both the beam model and clean component list positions, and are related through the given rotation matrix. The centre of the map is defined as pixel \((x,y) = (0,0)\). The mixing of \(Q\) and \(U\) is analytically identical, but of opposite sign, to that in Equation \ref{QUmixing}. The factor of \(2\) in the exponential originates from degenerate solutions for rotations of \(\pm180n\)~degrees. 

\begin{figure}[!hb]
\centering
    \resizebox{77mm}{!}{\includegraphics[height=60mm,angle=0,clip=true,trim=0cm 0cm 0cm 0cm]{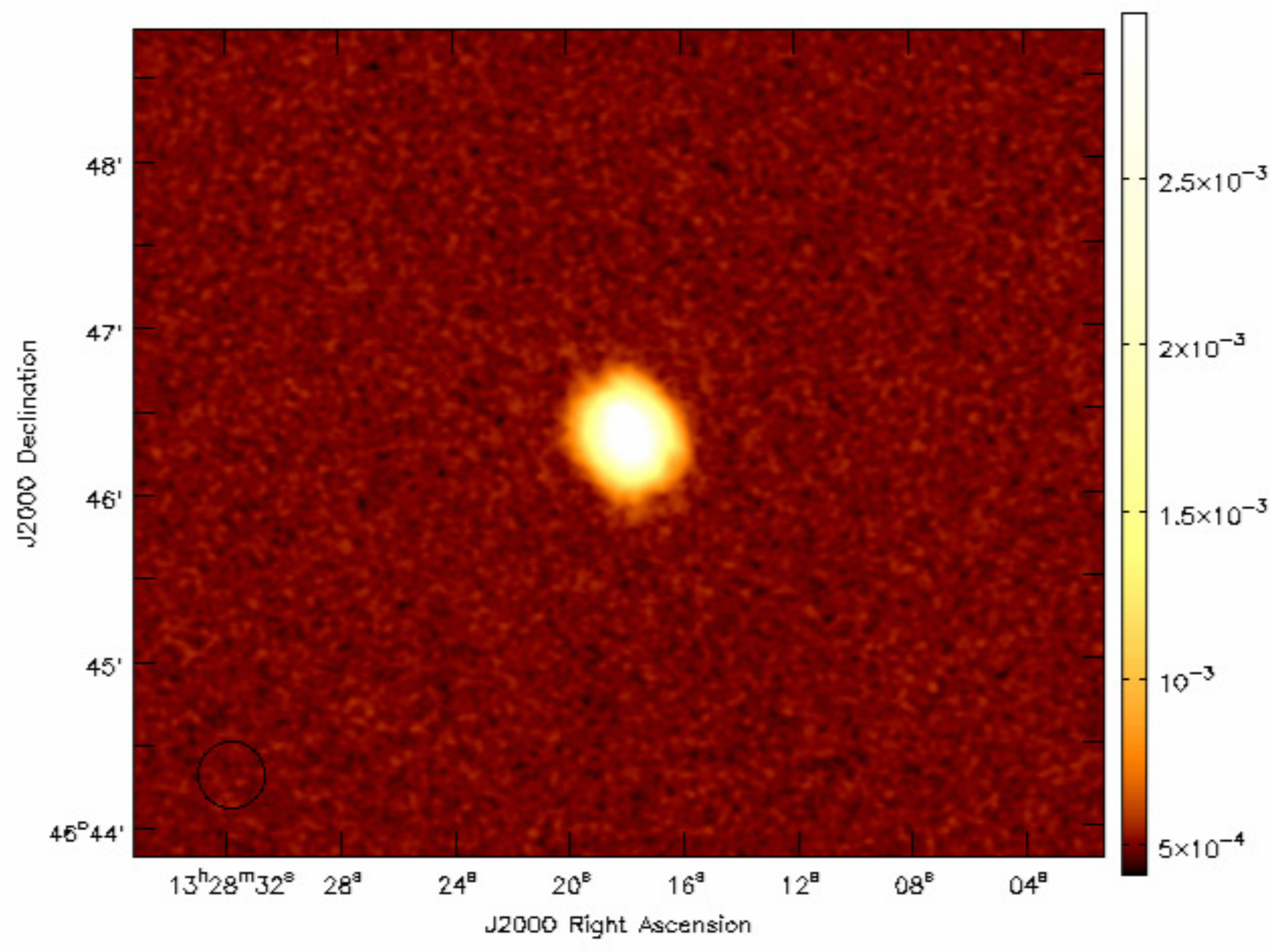}}
    \resizebox{77mm}{!}{\includegraphics[height=60mm,angle=0,clip=true,trim=0cm 0cm 0cm 0cm]{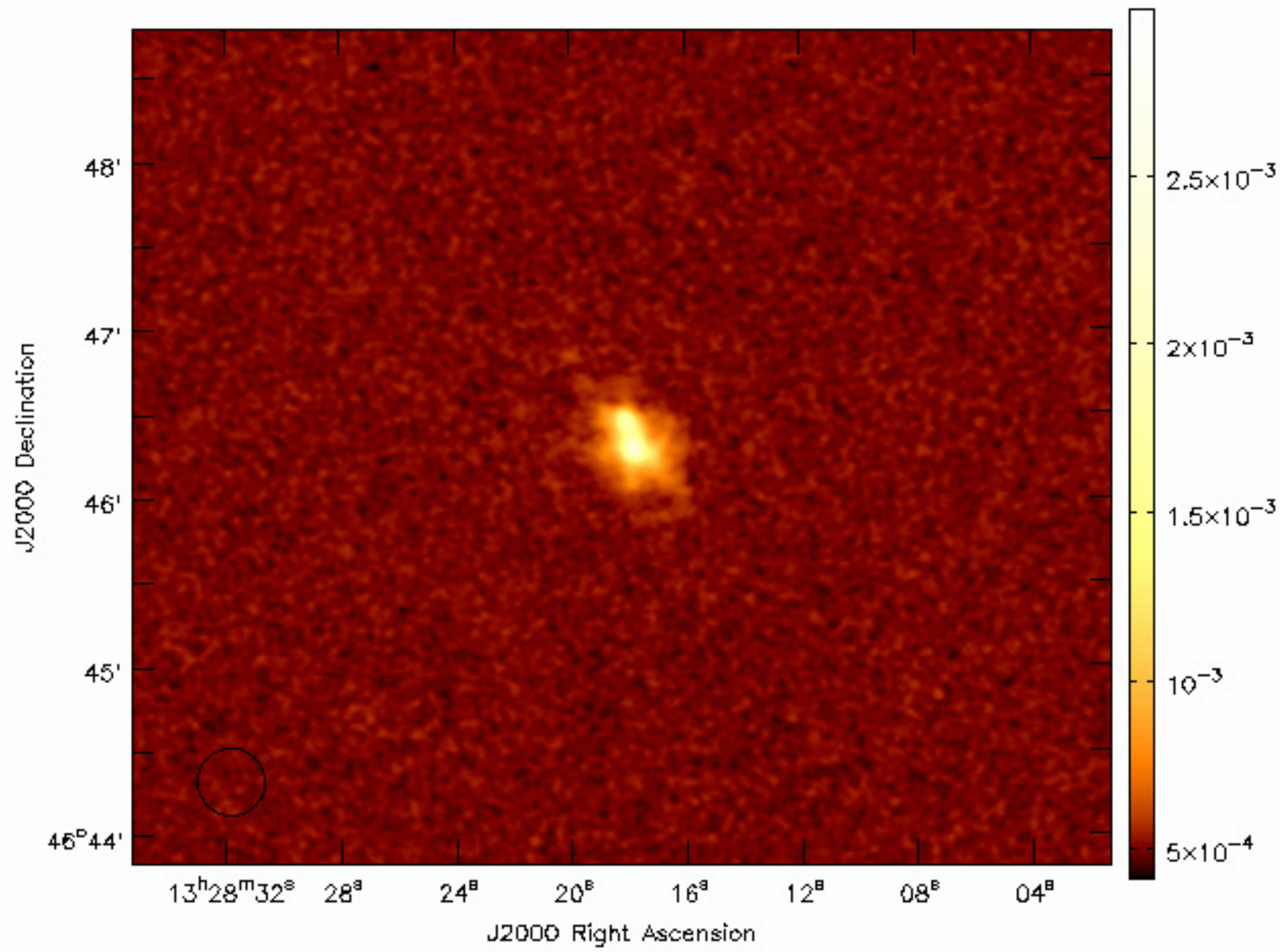}}
  \vspace{0pt} \\
    \resizebox{77mm}{!}{\includegraphics[height=60mm,angle=0,clip=true,trim=0cm 0cm 0cm 0cm]{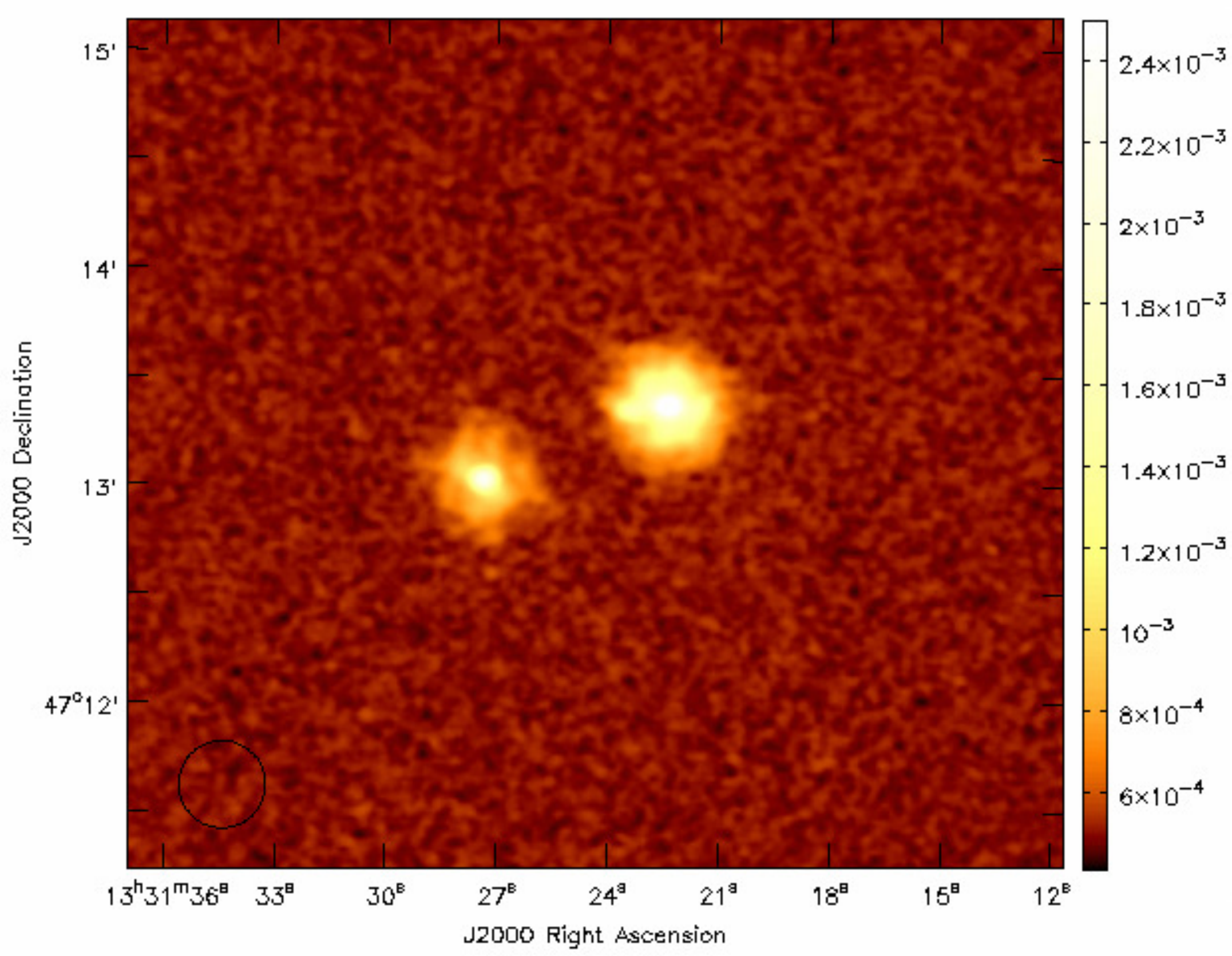}}
    \resizebox{77mm}{!}{\includegraphics[height=60mm,angle=0,clip=true,trim=0cm 0cm 0cm 0cm]{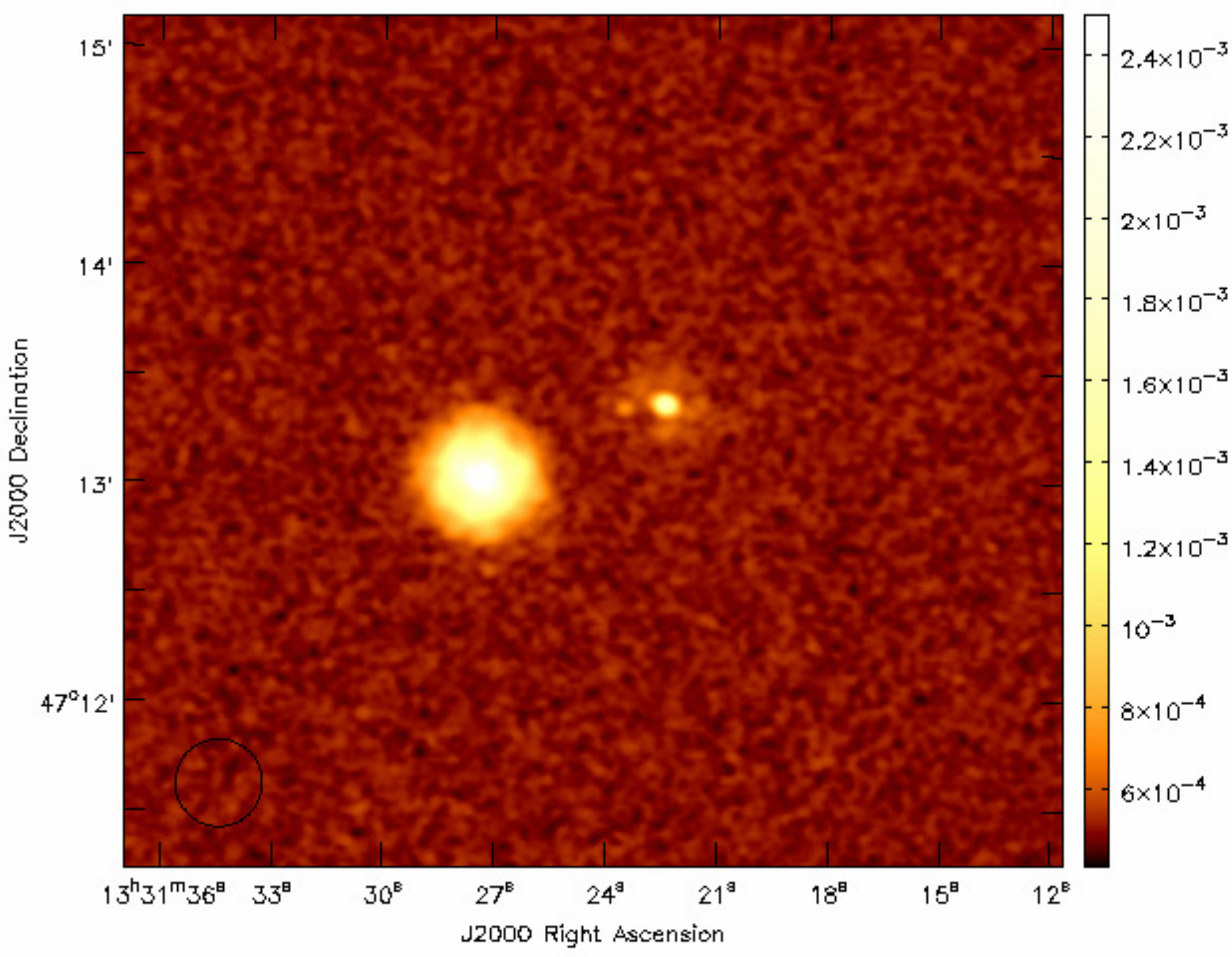}}
  \vspace{0pt} \\
  \caption[Polarised sources before and after corrections for off-axis effects]{\small{Images of the band-averaged polarised intensity both before (left) and after (right) the corrections for direction-dependent instrumental polarisation effects. Images are shown for two individual sets of sources as displayed to the top and bottom. These sources are located \(25.4^{\prime}\) and \(15.3^{\prime}\) off-axis respectively. One source (bottom) has its polarised intensity modified in an unusual way (see Section \ref{chap6:discussion}). The pseudo-colour scales are in units of Jy~beam\(^{-1}\), and the scales are identical for the before and after image of each source.}}
  \label{sourcesbeforeandafter}
  \vspace{0pt}
\end{figure}

The response is calculated for each chunk of approximately constant parallactic angle throughout the full-synthesis, and the Fourier Transform of the scaled clean components is then subtracted from the data. The entire \(uv\)-data is then re-imaged. This method should have the advantage that the associated sidelobes of the spurious instrumental polarisation are also removed from the final images, thereby increasing the dynamic range.

\begin{figure}[htb]
\centering
    \resizebox{115mm}{!}{\includegraphics[height=130mm,angle=0,clip=true,trim=0cm 0cm 0cm 0cm]{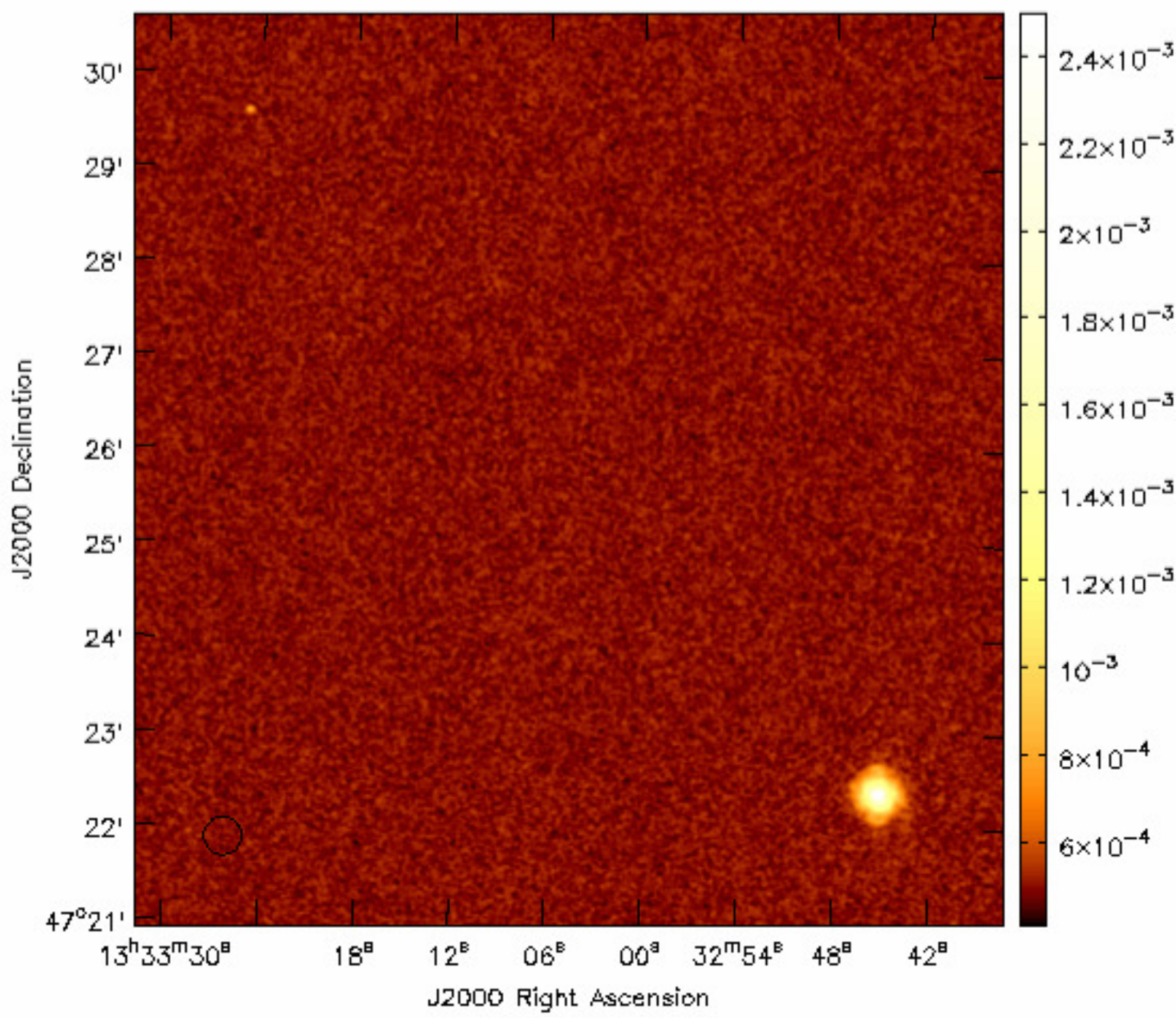}}
    \resizebox{115mm}{!}{\includegraphics[height=130mm,angle=0,clip=true,trim=0cm 0cm 0cm 0cm]{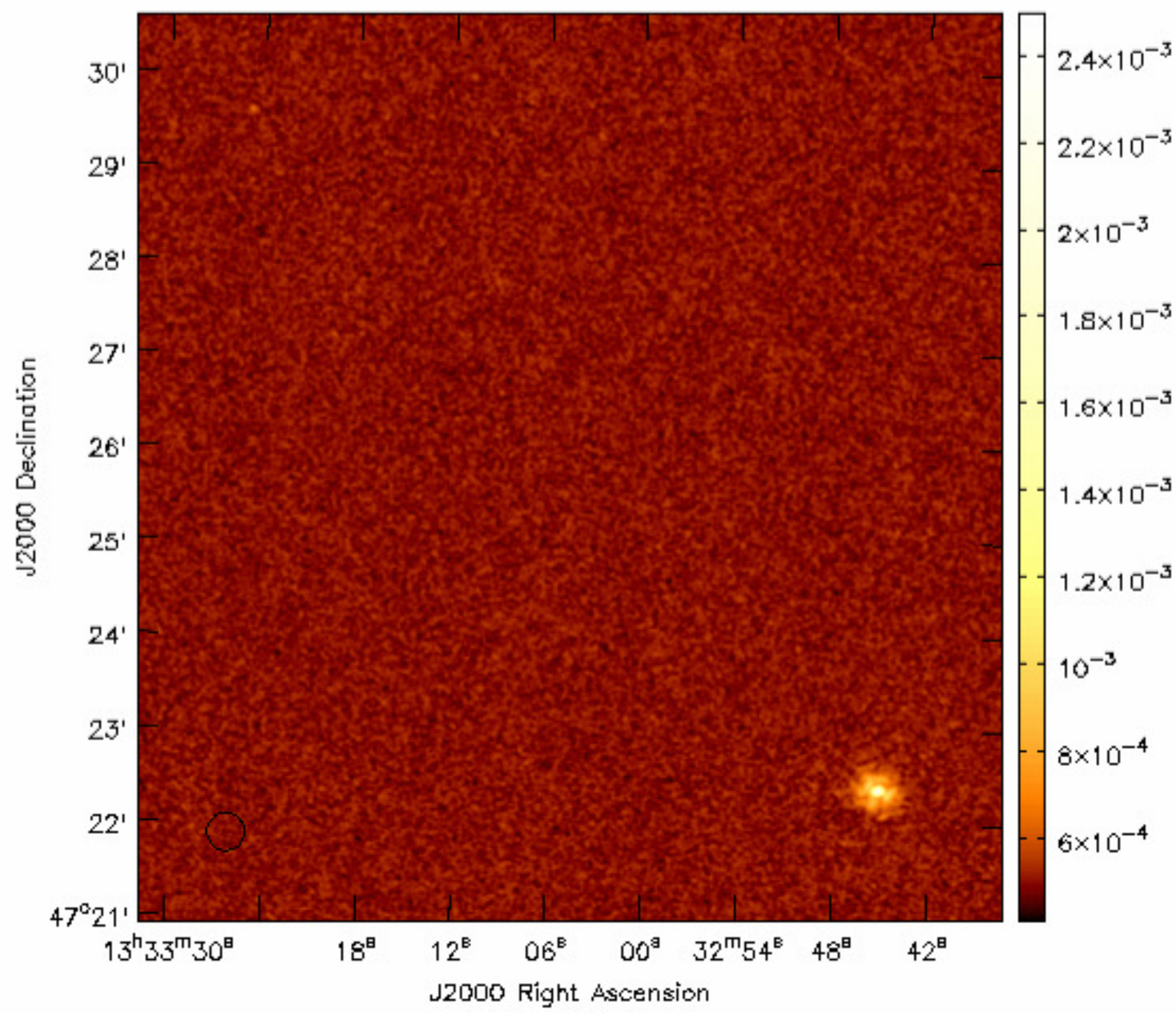}}
  \vspace{0pt} \\
  \caption[More polarised sources before and after corrections for off-axis effects]{\small{Similar to Fig.~\ref{sourcesbeforeandafter}, showing a map of the band-averaged polarised intensity both before (top) and after (bottom) the corrections for direction-dependent instrumental polarisation effects. Two sources -- one in the NE and another in the SW -- show reduced wide-field effects. The pseudo-colour scales are in units of Jy~beam\(^{-1}\), and the scales are identical for the before and after image. The sources are located \(14.1^{\prime}\) off-axis.}}
  \label{sourcesbeforeandafter1}
  \vspace{0pt}
\end{figure}

\begin{figure}[htb]
\centering
    \resizebox{119mm}{!}{\includegraphics[height=130mm,angle=0,clip=true,trim=0cm 0cm 0cm 0cm]{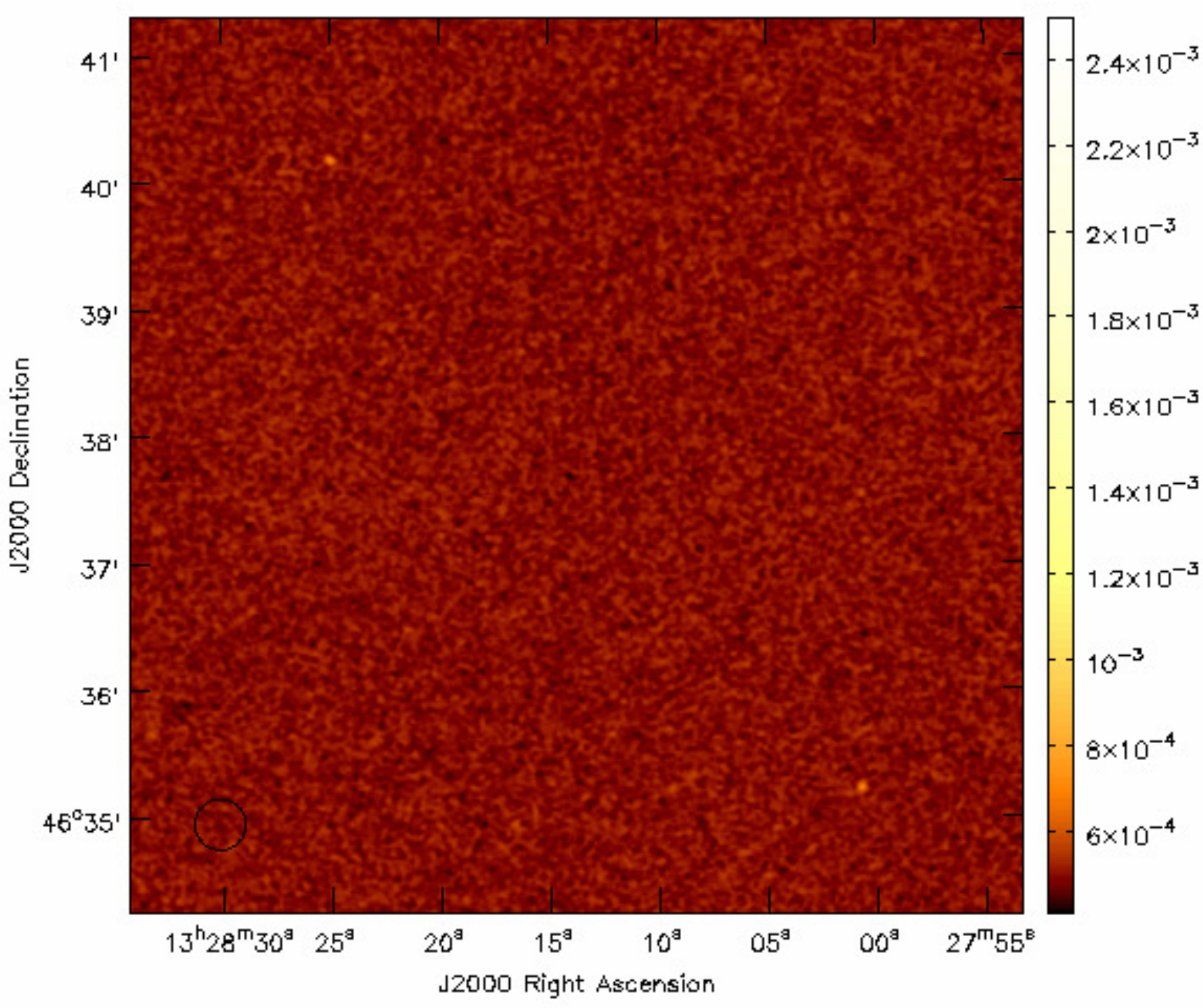}}
    \resizebox{119mm}{!}{\includegraphics[height=130mm,angle=0,clip=true,trim=0cm 0cm 0cm 0cm]{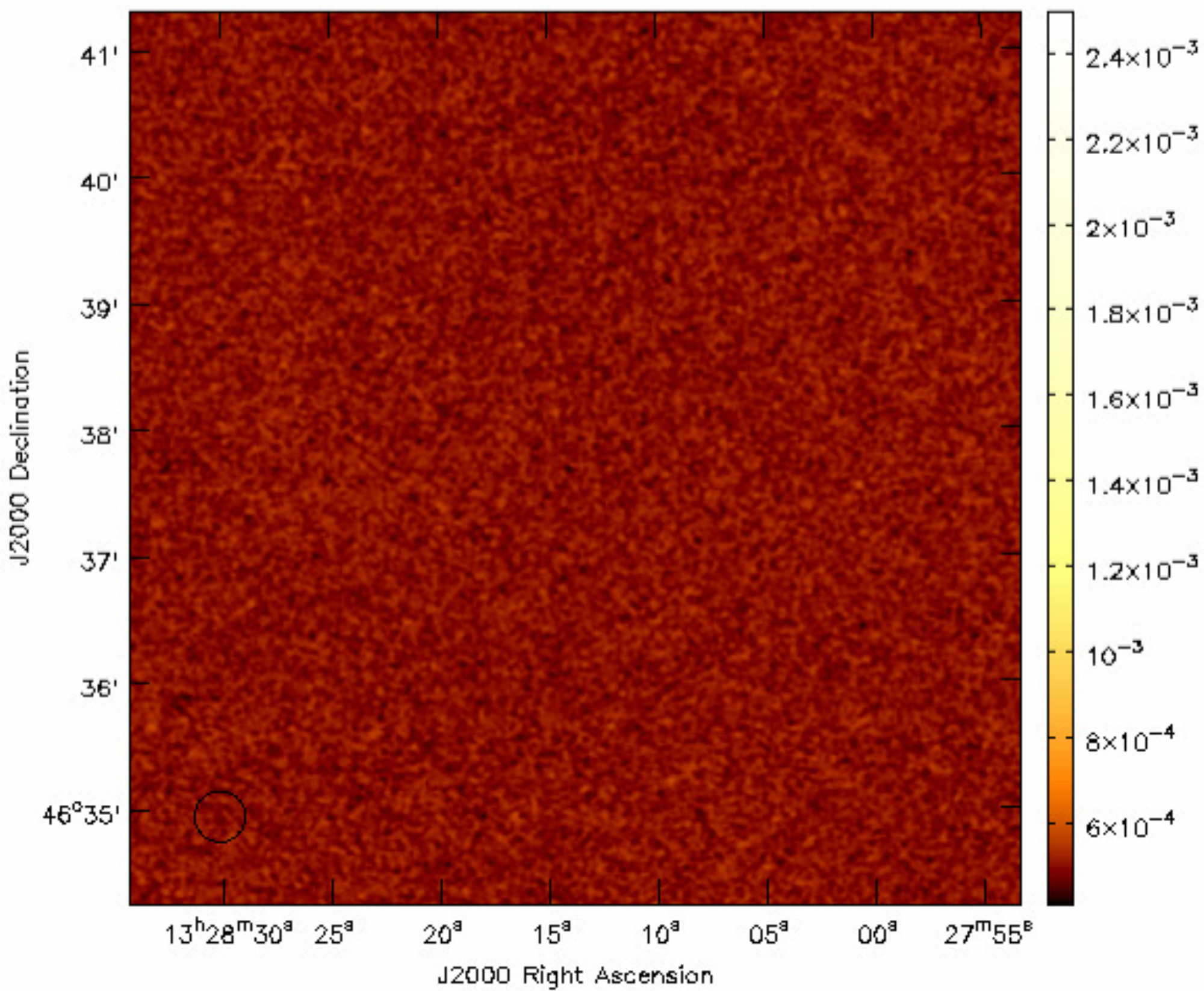}}
  \vspace{0pt} \\
  \caption[Even more polarised sources before and after corrections for off-axis effects]{\small{Similar to Fig.~\ref{sourcesbeforeandafter}, showing a map of the band-averaged polarised intensity both before (top) and after (bottom) the corrections for direction-dependent instrumental polarisation effects. Two sources -- one in the NE and another in the SW -- are removed completely from the map. The pseudo-colour scales are in units of Jy~beam\(^{-1}\), and the scales are identical for the before and after image. The sources are located \(34.3^{\prime}\) off-axis.}}
  \label{sourcesbeforeandafter2}
  \vspace{0pt}
\end{figure}

The impact of the corrections on selected off-axis sources is shown in Figs.~\ref{sourcesbeforeandafter} to \ref{sourcesbeforeandafter2}. The corrections shown here allowed each chunk to have a maximum parallactic angle variation of \(\Delta\chi=10^\circ\).

\clearpage
\section{A New Technique for Polarisation Angle Calibration}
\label{chap6:anewtechniqueforpacalibration}
\subsection{Limitations of Current Calibration Methods}
After solving for the instrumental leakage during polarisation calibration, the absolute value of the phase offset between \(R\) and \(L\) is left as an unconstrained degree of freedom. This typically needs to be corrected by observing a source with a known EVPA. The calibrators 3C138 and 3C286 are generally used for this purpose, with the RM and intrinsic EVPA measured at higher frequencies being used to interpolate to lower frequencies. This interpolation limits the quality of EVPA calibration, with errors in these \emph{a priori} values propagating into measurements of the RM from a target source. These errors are enhanced as the RM of 3C286 is typically assumed to be exactly equal to 0~rad~m$^{-2}$ \citep{2013ApJS..206...16P}, with other calibrators being measured relative to 3C286, and possibly adding a small systematic offset to a typically calibrated RM value. Consequently, no source on the sky has had it's RM directly measured using circular feeds; as is typical of interferometry, all RM measurements with circular feeds are \emph{relative} to some value from a calibrator source. 

This is particularly problematic for wide-band spectropolarimetry at low frequencies, as the assumed EVPA and RM of a calibrator source will be used to calibrate the source of interest. Even a small offset between the assumed and actual values at high frequencies can lead to a large difference at low frequencies, as the Faraday rotation scales with $\lambda^2$. This is possibly further exacerbated as different emitting regions within the source are possibly being probed at lower frequencies. In an illustrative example of the extent of this issue, consider an EVPA calibrator that is naively assumed to be well described by a single RM, but actually consists of multiple components at different Faraday depths. The single RM is then used to calibrate an observation of a target source, which is actually fundamentally described by a single RM. The flawed assumptions made about the EVPA calibrator will propagate into the target field, and RM Synthesis of the `calibrated' target will erroneously yield multiple Faraday components along the line of sight.

Furthermore, ionospheric variations and source variability also have an impact on the current techniques used for EVPA calibration. There are also a low number of sources with both a known, and a stable, intrinsic EVPA and RM at lower frequencies -- where depolarisation effects are significant. Overcoming this currently involves active searches for sources with `low variability' -- although these calibrator sources are rarely frequently monitored. Even in cases where the calibrators are monitored for variability, the variation is again only determined relative to yet another EVPA calibrator. It is clear that polarisation angle calibration at low frequencies would be substantially improved if we could therefore manufacture a polarised source in the sky with well-defined and stable properties.

\subsection{Manufacturing a Polarised Source}
We here suggest that having a suitable model of the wide-field response of an interferometer allows for EVPA calibration to be applied using the spurious off-axis polarisation from any bright \emph{unpolarised} source that is offset from the phase-centre. As the calibrator's `polarisation' is a consequence of the antenna, such a correction process is independent of ionospheric Faraday rotation. The process removes the need for calibration relative to any reference source, and instead makes the calibration relative only to the interferometer's primary beam. In this way, direction-dependent instrumental polarisation can become advantageous.

We now test this method using the \textsc{aips} package, although the technique could be extended for use in other packages, such as \textsc{casa} or \textsc{miriad}, if desired. For the specific example of the GMRT, the wide-field response is both frequency-independent and oriented radially. In such a case, in order to correct the data, the off-axis mixing of \(Q\)/\(U\) as described in Equation \ref{QUmixing} has to be taken into account. We assume that the data has already been calibrated for the effects of the complex gains, and for the instrumental leakages using e.g.\ \textsc{pcal}. During the standard step for polarisation angle calibration, the correction to be enacted in the \textsc{aips} task \textsc{clcor} is given by,
\begin{equation}
\textsc{clcorprm(1)} = 2\chi-180-\psi_{RL} \\,
\label{PAoffaxiscorr}
\end{equation}
where \(\chi\) is the parallactic angle of the off-axis scan of the unpolarised source, and \(\psi_{RL} = \psi_{LR^{\ast}}\) is the \(RL\)-phase as either output by task \textsc{rldif} or determined from the integrated \(Q\) and \(U\) values in the images such that,
\begin{equation}
\psi_{RL} = 2\phi = \arctan\left( \frac{\Sigma U}{\Sigma Q} \right) \\.
\label{PAoffaxiscorr2}
\end{equation}

Following calibration the measured EVPA of the off-axis source is given by \(\phi\), while the EVPA of the source in the antenna frame (i.e.\ independent of parallactic angle effects), \(\phi^{\prime}\), is given by,
\begin{equation}
\phi^{\prime} = \phi - \chi \\.
\label{PAoffaxiscorr4}
\end{equation}
This can be proven by taking Equation \ref{QUmixing} and substituting in \(Q = p_{0}\cos2\phi\) and \(U = p_{0}\sin2\phi\), which yields,
\begin{equation}
(Q^{\prime} + iU^{\prime}) = p_{0} \left\{ \cos(2\phi-2\chi) + i\sin(2\phi - 2\chi) \right\} \\.
\label{PAoffaxiscorr3}
\end{equation}

\subsection{Initial Tests on 325~MHz GMRT data}
\begin{figure}[!htb]
\centering
    \resizebox{130mm}{!}{\includegraphics[height=130mm,angle=0,clip=true,trim=0cm 0cm 0cm 0cm]{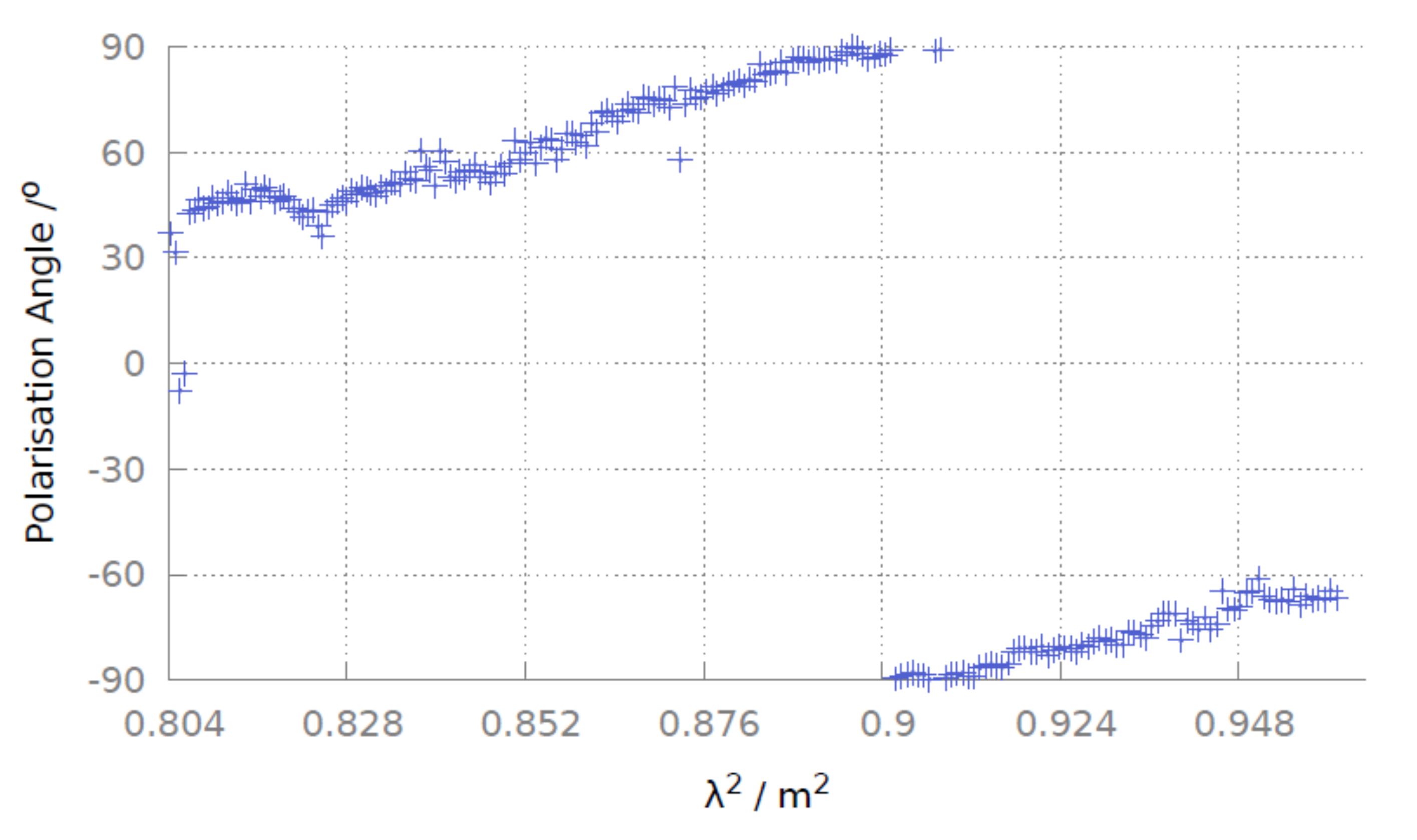}}
  \vspace{0pt} \\
  \caption[The EVPA versus squared-wavelength for B1937+21 following calibration using the direction-dependent response of the GMRT at 325~MHz]{\small{A plot of the EVPA as a function of \(\lambda^2\) for B1937+21 at 325~MHz, following polarisation angle calibration using an unpolarised source.}}
  \label{B1937+21}
  \vspace{0pt}
\end{figure}

We used the proposed technique to manufacture a polarised source using the off-axis response of the GMRT. We then used the technique to apply a polarisation angle calibration to GMRT data at 325~MHz. Our polarisation angle calibration was done using 3C48, which is unpolarised at these frequencies, and was located $80^{\prime}$ away from the phase-centre. The calibrated data were then used to determine the RM of pulsar B1937+21. The corrections yield a RM=\(+8.2\pm0.4\)~rad~m\(^{-2}\) for B1937+21 at 325~MHz -- in agreement with previous measurements at this frequency of \(+8.5\pm0.5\)~rad~m\(^{-2}\) \citep{2008A&A...489...69B}. The variations in the EVPA as a function of \(\lambda^2\) are shown for B1937+21 at 325~MHz in Fig.~\ref{B1937+21}.

\subsection{Extending to Linear Feeds}
The technique could also be extended for use with linear feeds, albeit in a more complicated form. One possible implementation could use the difference between the measured polarisation angles from the off-axis unpolarised calibrator, and what is expected from the beam model. This could then be used to iteratively refine the polarisation angle calibration. Using the known direction-dependent response of an interferometer to allow for calibration with an unpolarised source may be particularly useful for facilities such as the Australian Square Kilometre Array Pathfinder (ASKAP) \citep{2008ExA....22..151J}, the LOw-Frequency ARray (LOFAR) \citep{2013A&A...556A...2V}, and the Murchison Widefield Array (MWA) \citep{2013PASA...30....7T} -- where the science requires accurate determination of source RMs. At the very least, the technique can be easily used for verification of the quality of calibration. The technique presented here is limited only by knowledge of the beam model. The caveat is that the interferometer that is being calibrated must have significant direction-dependent instrumental polarisation.

\section{Discussion}
\label{chap6:discussion}
\subsection{Off-Axis Observing Strategy at the GMRT}
\label{discusssubsection}
The instantaneous \(Q\)/\(U\) beams deduced at 325~MHz and 610~MHz are compatible with the basic model presented in Section~\ref{intro}, as the direction-dependent effects appear to scale with the FWHM of the Stokes \(I\) beam -- implying that the beam squash is a result of interaction with the curved reflecting mesh, and not a consequence of the feed itself. 

The constraints placed on direction-dependent effects across the GMRT beam show that the snapshot 610~MHz instrumental response increases to \(\sim10\)\% at the half-power point, and increases rapidly to \(>30\)\% beyond 10\% of the Stokes \(I\) beam. The measured beam squash suggests that mosaicing snapshot observations would not be sufficient to overcome off-axis effects at the GMRT, as the half-power point snapshot \(P\) of \(\sim10\)\% at 610~MHz is greater than the expected median source polarisation of \(\sim2.5\)\% at this frequency. Nevertheless, \citet{farnesMexico} shows that this instantaneous instrumental polarisation can average down substantially over a large range in parallactic angle. The 325~MHz snapshot instrumental polarisation is similar to that at 610~MHz, and increases to \(\sim10\)\% at the half-power point, and to \(\sim30\)\% at 10\% of the Stokes \(I\) primary beam.

Consequently, full-track observations should be used to reduce direction-dependent effects at the GMRT. The use of mosaicing would reduce these off-axis effects further by an additional factor of \(\sim2\). As a conservative upper limit for observations taken over a large \(\chi\)-track, the instrumental polarisation at 610~MHz is at most \(\sim2.5\)\% at the half-power point, increasing steadily to \(>20\)\% beyond 10\% of the Stokes \(I\) beam.  Previous analyses of the 150~MHz GMRT beam are also possibly consistent with the off-axis response averaging down as a function of \(\chi\), with an instrumental polarisation of \(\sim2.5\)\% at the half-power point for a full-track observation \citep{2009MNRAS.399..181P}. However, no measurements of the instantaneous 150~MHz beam are currently available. Given the previously discussed scaling of direction-dependent effects with frequency, it is likely that the 150~MHz instantaneous beam is also simply a scaled version of the 610~MHz beam model shown in Fig.~\ref{allbeam}. 

\subsection{Towards a Unified Beam model}
Although not derived here, the scaling of the beam with frequency that is discussed in Section~\ref{discusssubsection} should in principle allow for the beam to be parameterised by a single mathematical function across all observing bands. This leads to the possibility that the GMRT Stokes \(I\) beam can also be defined by a single model for all frequencies, similar to the \(\cos^6(c \times n \times r)\) derived for the WSRT \citep{2008A&A...479..903P}. Variations in the Stokes \(I\) beam-width at the GMRT are expected, and the FWHM at 610~MHz is known to range from \(42.3^{\prime}\) to \(47.2^{\prime}\) in elevation, and from \(42.3^{\prime}\) to \(47.0^{\prime}\) in azimuth \citetext{Nimisha Kantharia: priv.\ comm.}. However, the beam results suggest that the 610~MHz primary beam has a squircle-like geometry, with the FWHM increased by \(\sim7.5\)\% along the diagonal axes -- coincident with the location of the feed support legs. Peeling is often required at the GMRT at this frequency, and may be a result of time-variable direction-dependent gains caused by the Stokes \(I\) beam rotating on the sky with varying \(\chi\). It would be interesting to employ more refined beam models in \(a\)-projection algorithms, which are able to remove such effects \citep{2008A&A...487..419B}.

\subsection{Circular Polarisation Effects}
The classical model of beam squint suggests that for observations of circular polarisation with a parabolic reflector and a feed perfectly aligned with the vertex of the paraboloid, there is perfect circular symmetry and therefore no beam structure in \(V\). However, if the feed deviates slightly from the vertex along the \(x\)-direction, then the two circularly polarised beams will point in slightly different directions along the \(y\)-direction \citep{2002ASPC.proc..278S}. This suggests that spurious \(V\) emission could be more significant in azimuth at the GMRT, as the rotating feed turret (which rotates about the elevation axis) may not lock exactly into place and point directly towards the vertex of the dish. Indeed, the Stokes \(V\) beam measured here is increased marginally in azimuth compared to elevation, by a magnitude of \(\sim1.5\)\%. It is possible the Stokes \(V\) beam therefore changes with each rotation of the feed turret.

\subsection{Correcting $uv$-data}
The polarisation beam corrections implemented in Section \ref{chap6:correctingresponse} are novel in that they directly correct the \(uv\)-data using a Stokes \(I\) model of an observed field. These corrections have been used to successfully correct for direction-dependent instrumental polarisation -- although by far less than had been hoped and with a number of limitations. The spurious linear polarisation is typically reduced by a factor of \(\sim\)2. Although useful, the beam corrections work best at correcting instrumental effects beyond the half-power point. The computational expense is therefore unlikely to be worthwhile, particularly when an equivalent or better reduction of the instrumental effects could be achieved via the combination of large parallactic angle ranges and mosaicing. Nevertheless, the corrections could be improved by implementing a cut-off for the Stokes \(I\) clean components, as noise and low-level artefacts that are cleaned in Stokes \(I\) currently get added into the linear polarisation images. Additionally some sources increase in \(P\) as shown in Fig.~\ref{sourcesbeforeandafter}. This can be expected depending on the precise combination of source and instrumental polarisation, but may also suggest that the beam model is not sufficiently accurate. The latter is likely as the direction-dependent corrections implemented here reduce the instrumental polarisation most effectively at distances \(>30^{\prime}\) from the phase-centre i.e.\ where the corrections are of a larger magnitude. Limitations to the corrections may result from the decrease in signal-to-noise of rasters near to the phase-centre, significant beam variation between individual antennas, the coarse sampling used, or because of assumptions of \(\chi\) constancy in each snapshot. In the corrections presented in this report, each chunk has \(\Delta\chi < 10^{\circ} \), and no attempt to test the effects of different \(\Delta\chi\) has been made. Nevertheless, reasonable wide-field polarimetry is possible at the GMRT via the use of full \(\chi\)-tracks. Such direction-dependent corrections will be most useful in efforts to retrieve the intrinsic EVPA of sources, which are clearly corrupted by the instrumental beam \citep{farnesMexico}.

\subsection{Summary and Future Work}
New data are needed to improve the beam model and consequently the corrections. An attempt to monitor the beam out to the first sidelobe, with more axes, and less sparsely-sampled rasters would significantly enhance the model. The recommendation would be that tests are performed, the feed turret rotated, and the tests repeated in order to rule out epoch-dependent variations due to locking of the feed turret mechanism. This would also allow for checks on the observed change in sign along the Elevation-X axis. Such observations, if taken for all GMRT bands, may provide the possibility of describing the beam with a unified model which is a function of frequency.

Knowledge of a telescope's beam is extremely important, as demonstrated by the ability to calibrate the polarisation angle using an unpolarised source, when the direction-dependent response is known. This allows for corrections that are independent of ionospheric Faraday rotation effects and a number of other low-level systematics. Importantly, this technique can be extended down to arbitrarily low frequencies. Improving the beam model at the GMRT is one of the most important endeavours that could currently be carried out to refine the instrumental response, and follow-up observations should be strongly considered. Such observations may lead to improvements in the pointing model and dynamic range at all frequencies.

\addcontentsline{toc}{section}{References} 


\addcontentsline{toc}{section}{Acknowledgements} 
\section*{Acknowledgements}
I am exceptionally grateful to Dave Green, for providing detailed feedback on the manuscript, Nimisha Kantharia, for spending extensive time obtaining the high-quality test observations used within this report, and also to Anna Scaife and Russ Taylor for many insightful conversations and useful comments on the manuscript. We thank the staff of the GMRT that made these observations possible. The GMRT is run by the National Centre for Radio Astrophysics of the Tata Institute of Fundamental Research. J.S.F. has been supported by the Science and Technology Facilities Council, UK.






\small
\begin{thebibliography}{}
 \bibitem[Bhatnagar et al.(2008)Bhatnagar et al.]{2008A&A...487..419B} 
Bhatnagar S., Cornwell T.~J., Golap K., Uson J.~M., 2008, A\&A, 487, 419. 
 \bibitem[Brentjens(2008)Brentjens]{2008A&A...489...69B} 
Brentjens M.~A., 2008, A\&A, 489, 69. 
 \bibitem[Brentjens \& de Bruyn(2005)Brentjens \& de Bruyn]{2005A&A...441.1217B} 
Brentjens M.~A., de Bruyn A.~G., 2005, A\&A, 441, 1217. 
\bibitem[Carretti et al.(2004)Carretti]{caretti2004}
  Carretti E., et al., 2004, A\&A, 420, 437.
 \bibitem[Cotton(1994)Cotton]{CottonAIPSMemo} 
Cotton W.~D., AIPS Memo \#86, ``Widefield Polarisation Correction of VLA Snapshot Images at 1.4~GHz'', 1994.
 \bibitem[Farnes(2012)Farnes]{Farnesthesis} 
Farnes J. S., 2012, ``Polarimetric Observations at Low Radio Frequencies'', PhD thesis, University of Cambridge.
 \bibitem[Farnes et al.(2013)Farnes et al.]{farnesMexico} 
Farnes, J. S., Green, D. A., Kantharia, N. G., 2013, in `Magnetic Fields in the Universe:
from Laboratory and Stars to Primordial Structures', edited by A. Esquivel,
J. Franco, G. García-Segura, E. M. de Gouveia Dal Pino, M. Krause, A. Lazarian.
\& A. Santillan, in press (arXiv:1309.4646).
 \bibitem[Farnes et al.(2014)Farnes et al.]{farnesMNRAS} 
Farnes, J. S., Green, D. A., Kantharia, N. G., 2014, MNRAS, 437, 3236.
 \bibitem[Farnsworth et al.(2011)Farnsworth, Rudnick, \& Brown]{2011AJ....141..191F} 
Farnsworth D., Rudnick L., Brown S., 2011, AJ, 141, 191.
 \bibitem[Garn(2009)Garn]{2009PhDT.........3G} 
Garn T.~S., 2009, ``610~MHz observations of galaxy evolution'', PhD thesis, University of Cambridge.
 \bibitem[George et al.(2011)George, Stil, \& Keller]{2011arXiv1106.5362G} 
George S.~J., Stil J.~M., Keller B.~W., 2012, PASA, 29, 214. 
 \bibitem[Heiles(1999)Heiles]{HeilesAreciboMemo} 
Heiles C., Arecibo Observatory Technical Memo AOTM 99-02, ``The LBW Feed: Pointing Accuracy, Beamwidth, Beam Squint, Beam Squash', 1999.
 \bibitem[Heiles(2002)Heiles]{2002ASPC.proc..278S} 
Heiles C., ``Single-Dish Radio Astronomy: Techniques and Applications'', ASP Conference Proceedings, 278, 2002.
 \bibitem[Heiles et al.(2003)Heiles et al.]{HeilesGBTMemo} 
Heiles C., Robishaw T., Troland T., Anish Roshi D., ``GBT Commissioning Memo \#23'', 2003.
\bibitem[Johnston et al.(2008)]{2008ExA....22..151J} 
Johnston, S., Taylor, R., Bailes, M., et al., 2008, Exp. Astron., 22, 151. 
 \bibitem[Pen et al.(2009)Pen et al.]{2009MNRAS.399..181P} 
Pen U.-L., Chang T.-C., Hirata C.~M., Peterson J.~B., Roy J., Gupta Y., Odegova J., Sigurdson K., 2009, MNRAS, 399, 181.
\bibitem[\protect\citeauthoryear{Perley \& Butler}{2013}]{2013ApJS..206...16P} 
Perley R.~A., Butler B.~J., 2013, ApJS, 206, 16.
 \bibitem[Popping \& Braun(2008)Popping \& Braun]{2008A&A...479..903P} 
Popping A., Braun R., 2008, A\&A, 479, 903. 
 \bibitem[Simmons \& Stewart(1985)Simmons \& Stewart]{1985A&A...142..100S} 
Simmons J.~F.~L., Stewart B.~G., 1985, A\&A, 142, 100. 
 \bibitem[Taylor et al.(2009)Taylor, Stil, \& Sunstrum]{2009ApJ...702.1230T} 
Taylor A.~R., Stil J.~M., Sunstrum C., 2009, ApJ, 702, 1230.
 \bibitem[Tinbergen(1996)Tinbergen]{1996aspo.book.....T} 
Tinbergen J., ``Astronomical Polarimetry'', Cambridge University Press, 1996.
\bibitem[Tingay et al.(2013)]{2013PASA...30....7T} 
Tingay, S.~J., Goeke, R., Bowman, J.~D., et al., 2013, PASA, 30, 7. 
\bibitem[van Haarlem et al.(2013)]{2013A&A...556A...2V}
van Haarlem, M.~P., et al., 2013, A\&A, 556, 2.
 \bibitem[Wardle \& Kronberg(1974)Wardle \& Kronberg]{1974ApJ...194..249W} 
Wardle J.~F.~C., Kronberg P.~P., 1974, ApJ, 194, 249. 
\end{thebibliography}
\end{document}